\documentclass[11pt]{article}

\usepackage[margin=.85in]{geometry}
\usepackage{setspace}
\usepackage{amsmath,amssymb,amsthm}
\usepackage{natbib}
	\setcitestyle{aysep={}}
\usepackage{pdfsync}
\usepackage{latexsym}
\usepackage{amsthm}
\usepackage{multibib}
\usepackage{hyperref}
\usepackage{soul}
\usepackage{booktabs}
\usepackage{tabularx}
\usepackage{enumitem}

\usepackage{titlesec}
\usepackage{xcolor}

\definecolor{tealstrong}{RGB}{60,140,140}
\definecolor{purplestrong}{RGB}{140,100,160}
\usepackage{mathpazo}

\titleformat{\section}
  {\Large\bfseries}
  {\thesection}{0.5em}{}

\titlespacing*{\section}{0pt}{1.2em}{0.5em}
\titlespacing*{\subsection}{0pt}{1em}{0.3em}

\usepackage{tikz}
\usetikzlibrary{arrows.meta,fit,shapes,backgrounds,positioning,patterns,decorations.pathreplacing}
\usepackage{epigraph}
\usepackage{csquotes}

\usepackage{appendix}
\usepackage{lipsum}
\usepackage{enumitem}

\usepackage[font=small,labelfont=bf,labelsep=colon]{caption}

\newcommand{\E}{\mathbb{E}}

\begin{document}

\begin{flushleft}
{\bfseries\Large \textbf{What Is the Causal Effect of a Conversation?}} \\[0.15em]
{\bfseries\large \textbf{Estimands and inference in AI mediated conversations}} \\[0.75em]

{ \large Adriane Fresh\footnote{Interdisciplinary Data Science Program, and Department of Political Science, Duke University; adriane.fresh@duke.edu.}, Aiden Shin\footnote{Department of Political Science, Duke University; juneseok.shin@duke.edu.} \\ This draft: \today}
\end{flushleft}

\vspace{0.2in}
\hrule
\vspace{0.15in}

\begin{abstract}
\noindent
Political scientists increasingly use conversations as treatments.  Sometimes these conversations are conducted by humans: canvassers, interviewers, deliberative moderators, or peers.  Often and increasingly, however, they will be conducted by generative artificial intelligence (AI).  The methodological attraction is obvious.  AI systems make it possible to scale treatments that are interactive, responsive, and rich in content and real-world relevance.  But this same interactivity creates specific challenges for causal inference.  A respondent may be randomly assigned to a prompt, a persona, or a conversational condition, but the conversation that follows is not merely received by the respondent.  It is generated \emph{jointly} by the respondent and the conversational agent and is thus endogenous to who the respondent is.  Consequently, when the theoretical object of interest is the conversation itself---or particular messages, features, or other attributes of that conversation---randomization of assignment does not necessarily identify the causal quantity the researcher seeks to estimate.

This paper develops a potential outcomes framework for causal inference with conversations, with broad application but particular relevance to AI-mediated interaction. We distinguish among several causal objects, including assignment to a conversational condition, assignment to a conversational policy, opening messages, messages within a conversation, realized conversational features, and the full realized conversational event. Each corresponds to a distinct estimand and a different set of identifying assumptions. While some of these quantities are identified by standard randomized designs, others require additional assumptions or research designs, including sequential assumptions, representations of conversational histories, or explicit message-level intervention regimes. The framework clarifies these distinctions and provides a common language for defining, interpreting, and designing conversational experiments as conversations increasingly become objects of causal inquiry.
\end{abstract}

\vspace{0.15in}
\hrule
\vspace{0.2in}

\section{Motivation}

\noindent Social interaction occupies a central place in both theoretical and empirical accounts of political behavior \citep{Goffman:1983xe,Habermas:1984fs}.  Individuals do not form political beliefs and attitudes in isolation.  They develop, revise, and express them in interaction with others \citep{Lazarsfeld:1944fr,Huckfeldt:1995kw,Mutz:2006ka}.  A central mechanism through which this interaction is realized is the \emph{conversation}---a sequential, interactive linguistic exchange between two or more participants \citep{Sacks:1974so,Clark:1996nn}.

Conversations allow individuals to update factual beliefs, to calibrate their perceptions of what others think, to learn which positions are socially acceptable, to refine their own arguments in light of agreement or disagreement, and more.  They also frequently serve expressive and relational functions---individuals signal identity, establish group boundaries, construct shared meaning and negotiate status \citep{Searle:1992wl}.  From this perspective, it is natural that political science has long treated conversations as a critical causal force operating in the social world.

In light of their theoretical relevance, conversations are increasingly attractive causal objects for empirical scholars because they do something that simpler political interventions often cannot.  They \emph{respond}.  A leaflet, advertisement, vignette, newspaper headline or image can present information to a subject, but it cannot echo back the subject's stance, ask why the subject disagrees, reframe a claim in light of disagreement, or alter its tone when the subject becomes hostile.  A conversation, however, can.  Thus, a conversation more closely approximates the real-world environment---between people, via social media, and with generative artificial intelligence (AI)---in which politics operates.   For these reasons, conversations have become increasingly central to studies of persuasion, tolerance, deliberation, political disagreement, polarization and social media behavior \citep{Huckfeldt:1995kw,Pattie:2001wl,McClurg:2004ss,Pattie:2009nn,Broockman:2016mq,Pattie:2016yb,Pons:2018vv,Kalla:2020oc,Munger:2021du,Kalla:2022yz,Mikli:2022uv,Combs:2023ai,Argyle:2025zo,Lin:2025ga,Hackenberg:2025th,Walter:2025mc,Allamong:2025ffq,Allamong:2026jy}.  Moreover, the rapid proliferation of generative AI has made the scalability of conversational treatments much greater \citep{Altay:2022gd}.

At the same time, it is the responsive character of conversation that makes causal inference in such research difficult.  If a respondent is assigned to read a fixed piece of text, the treatment can be described with some precision.  By contrast, when a respondent is assigned to have a \emph{conversation} that begins with that text, the subsequent exchange is not fixed by the investigator alone.  It is generated jointly over time through interaction.  The interlocutor's contributions---in content, tone, length, timing and so forth---may vary across respondents, and the respondent's own conversational tendencies---curiosity, hostility, verbosity, resistance, humor, defensiveness, eloquence, and so on---help determine what is said next.    As a result, the conversation that is ultimately experienced is jointly produced rather than assigned (see e.g, \cite{Zhang:2021zq}).  In AI-mediated settings, these challenges do not fundamentally differ from those in human conversations, but they become more prevalent and will arguably become harder to ignore.

This aspect of conversations, we argue, has important implications for causal inference.  Random assignment to a conversational condition identifies the causal effect of assignment to that condition.  However, it does not, without additional assumptions, identify the effect of the realized conversation or of the conversational content that the assignment is intended to induce.  Even under randomization, variation in conversational exposure may be systematically related to respondent characteristics that also affect outcomes, complicating the interpretation of certain estimands.

To illustrate, consider the following brief example of a study conducted by a researcher interested in the effect of civility on attitudes towards democracy, an example we will return to later in this paper. Suppose a researcher is interested in civil political conversation and its relationship to attitudes towards the integrity of democratic institutions.\footnote{Along the lines of, but distinct from \cite{Allamong:2025ffq}.} Respondents are recruited into a study and randomly assigned to interact with an AI chatbot (backed by a large language model) about a political topic. Some respondents are assigned to a chatbot configured to engage in a civil and respectful manner, while others are assigned to a chatbot configured to engage in a more confrontational or uncivil manner. After the conversation, respondents are asked their attitudes about democratic institutions.

At first glance, the causal question appears straightforward---does a civil conversation increase certain perceptions of democratic institutions? But on closer examination, it is not actually so clear that the design recovers that causal effect.  To see this, consider two participants in the study---a 40 year old therapist and a 22 year old internet troll---both assigned to the ``incivil'' condition. The AI opens both conversations in the same way, but because of who these participants are, and how they engage with incivility, and thus respond to the chatbot, the conversations quickly diverge. In the first conversation, the therapist is able to quickly steer the conversation towards civility.\footnote{While the AI has a prompt that inclines it towards incivility, AI chatbots are designed to recapitulate patterns they have observed in their training data.  Consequently, they respond to the therapist’s efforts at de-escalation much as would a real person.} As a result, the actual level of incivility in the resulting conversation (e.g., the incivility score a third party who only observed the messages in the conversation might assign the conversation) is actually quite low.  In the second conversation, by contrast, the internet troll immediately levies an insult and the level of incivility quickly escalates.  Any independent observer would conclude the level of incivility in the conversation was off the charts.

Thus while both participants had been assigned to the same condition, and assigned via randomization, the actual content of \emph{the conversation} that they experienced was quite different.  The way it was different was as a result of pre-existing features of the participants, features which are likely to be correlated with the outcome of interest of the study, their perceptions of democratic institutions.  Even if we turn our attention to features of the conversation, or a given message, variants of this confounding potentially remain.

To deal with this nuance of interactive conversations, this paper provides a framework for defining and interpreting causal effects in conversational settings, where the treatment takes the form of a sequential, interactive process rather than a fixed intervention.  We show that defining causal effects in such settings requires distinguishing between distinct theoretical objects: (i) the generalized conversational condition, (ii) the opening message, (iii) the conversational \emph{policy} or regime that governs how interaction unfolds \citep{Graves:2000js}, (iv) an individual messages with a conversation, and (v) the realized conversational exposure that results from that interaction.  Each of these objects corresponds to a different causal estimand, and each requires different assumptions for interpretation.  Our paper defines those estimands and the relevant assumptions.  It then shows what is implied for research designs interested in these distinct objects.

In doing so, our framework makes transparent a tension in contemporary causal inference.  Much experimental work has focused on interventions that are discrete, well-defined, and tightly bounded in time and content.  Such treatments respect Holland's adage that without manipulation there is no identification \citep{Holland:1986ut}.  Consequently, one response to the difficulties posed by more complex interventions is to simply restrict attention to the causal effect of assignment---the \emph{intent-to-treat} (ITT) effect---which tends to be straightforwardly identified with standard research designs.  While internally valid, such estimands may be only loosely connected to a target treatment of theoretical interest \citep{Egami:2023jz}.\footnote{This is related to the motivating point made by \cite{Fong:2023um}.  In the context of textual treatments, scholars may be able to narrowly define the causal effect of a set of ordered tokens, but require different estimands and, or, additional assumptions in order to make causal claims about the latent features---tone, opinion or topic---that are the actual object of theoretical interest.  }  While some scholars emphasize their research questions in relation to real world interventions that could be adopted---e.g., so-called policy relevance---others focus more broadly on understanding \emph{why} the political world works as it does, which often moves beyond these neatly bounded interventions.  As researchers push treatments to become more realistic, the gap between the intervention and the theoretical object of interest can widen appreciably.  The approach taken here in this paper is not to retreat to narrowly defined estimands due to the ease of identification, but to clarify what causal questions can be posed in conversational settings and what assumptions are required to answer them.

\section{Contribution}

\noindent This paper relates to several strands of work in causal inference and computational social science.  First, it connects to a growing literature on causal inference with text, which considers language as a treatment, outcome, or mediator, and emphasizes the challenges that arise when linguistic objects are high-dimensional \citep{Roberts:2020ki,Grimmer:2022ih,Egami:2022ax,Feder:2022jb}.  A central insight from this work is that text often bundles multiple latent features that must be represented as some function of text, and interacts with individual's existing beliefs, making it difficult to isolate the causal effect of any particular component or attribute \citep{Dafoe:2018uv,Roberts:2020ki,Fong:2023um}.  Conversational settings, however, extend these issues because the representation problem is no longer static but dynamic: the researcher must approximate not only the features of a message, but the evolving interpretive context in which that message is understood.  This substantially expands the scope of the representation problem and requires care in the application of those approaches.

Second, the paper draws on work in sequential and dynamic causal inference, where treatments are administered over time and may depend on prior outcomes and histories \citep{Robbins:2986me,Murphy:2003du,Murphy:2005aa,Laber:2014ry}.  In such settings, the causal effect of an intervention is defined with respect to a regime governing the sequence of actions, rather than a single static treatment.  However, path-dependent counterfactuals can mean that interventions are not merely unobserved, but instead are undefined or ill-posed given the evolution of the interaction \citep{Slough:2023ee}.

The paper also relates to emerging work on conversational interaction, including studies of AI-mediated communication, which highlight the inherently interactive and co-produced nature of linguistic exchange \citep{Zhang:2020qp,Zhang:2021zq,Shen:2021yy,Tuan:2022hn}.\footnote{At a more granular level, recent work has examined the causal effect of individual word substitutions within static text by treating sentences as units and conditioning on observed context \citep{Wang:2019oo}.}  In particular, \cite{Zhang:2021zq} develops a framework in which the meaning and function of an utterance are defined by the distribution of conversational contexts in which it appears.  This perspective highlights that utterances are not fixed treatments, but relational objects whose interpretation depends on both prior interaction and anticipated continuation.  While this work provides a rich descriptive account of conversational dynamics, it stops short of defining causal estimands for conversational interventions, or specifying the assumptions under which such estimands can be identified.  The framework developed here builds on these insights by formalizing the causal objects that arise in conversational settings.

These conceptual distinctions matter because conversational interventions are increasingly being deployed experimentally.  These existing conversational studies typically randomize assignments to prompts, systems, policies, or conversational environments \citep{Argyle:2025zo,Lin:2025ga,Hackenberg:2025th,Walter:2025mc,Allamong:2025ffq}. These interventions are often well-defined and experimentally tractable. At the same time, the substantive theories motivating such studies frequently concern broader conversational objects; theoretical concepts like persuasion, deliberation, incivility, empathy, or interaction itself. The distinction is not a flaw in existing work. It instead reflects the fact that conversational settings contain multiple plausible and theoretically interesting causal objects. The framework developed here in this paper provides a common language for distinguishing among these objects, and it identifies which are directly manipulated by a design, and clarifies the assumptions required to move from one to another.

Lastly, the inferential challenges identified in this paper are related to, but distinct from, concerns in the literatures on causal mediation.  Work on mediation highlights the difficulty of interpreting causal effects when treatments operate through intermediate variables that are themselves affected by the treatment \citep{Keele:2011uz,Acharya:2016tt,Acharya:2018ww}.  In conversational settings, features such as incivility, empathy, correction, or perspective-taking can be thought of as mediators relative to assignment, but they need not be \emph{only} mediators for the purposes of theory.  In many applications, these features are themselves the causal objects of interest.\footnote{One person's mediator is another person's causal object of interest.  } Moreover, the path-dependent and multi-stage nature of conversation significantly expands the set of causal \emph{paths}. The problem, therefore, is not solely how to decompose the effect of assignment into mediated paths, but how to define and identify causal effects for features of a conversational process that may be, at least partially, generated after assignment.

The contribution of this paper is to bring these strands of work into a common framework tailored to conversational settings.  Existing research has made important progress in characterizing text as a causal object, in defining effects under sequential regimes, and in analyzing interactional data.  However, these literatures have largely treated their respective objects---texts, treatments, or interactions---in isolation.  By contrast, conversations combine these features: they are linguistically rich, extend over time, and are generated jointly.  As a result, the mapping between assignment, realized exposure, and outcomes is more complex than in standard settings.  This paper contributes by making that mapping explicit in order to clarify what can and cannot be learned about them under common research designs.

\section{The Basic Setup}\label{sec:setup}

\subsection{Scope}

\noindent To make tractable the definition of causal estimands in conversational settings, we focus on a narrow but still representative subset of conversations.\footnote{The definition of a conversation adopted here is intentionally narrow. Many politically relevant interactions involve more than two participants, asynchronous communication, branching discussion threads, or looser patterns of exchange than the turn-taking structure assumed in this paper. Examples include online comment sections, discussion forums, group deliberation, social media exchanges, and multi-party conversations. Whether such interactions should be treated as a single conversation, multiple linked conversations, or some other conversational object depends partly on the research question and the causal object of interest. The framework developed here is therefore best understood as a tractable starting point whose basic logic can be extended to more complex interactional settings.}  In particular, we consider interactions between two actors---a \emph{respondent} and an \emph{interlocutor}.  Because of our interest in AI-mediated settings, the interlocutor will frequently be referred to as an AI, though the framework extends naturally to human interlocutors as well (Section~\ref{sec:humans}).

The two actors exchange messages sequentially.  A minimal conversation consists of each actor providing at least one message, though the inferential challenges of interest become more apparent in longer exchanges ($>3$ total messages).  We therefore focus on multi-turn conversations in which each message may depend on the prior history of interaction.  We restrict attention to \emph{linguistic} conversations, such that messages can be represented as text.  Real-world conversations contain additional paralinguistic elements---eye contact, facial expressions, vocal intonation, volume, gesturing, and so forth---but even the textual component alone introduces substantial complexity (see e.g., \cite{Wang:2017dd,VanZant:2020ps}).\footnote{Even deeper than paralinguistic elements is the \emph{absence} of text.  As Dutch author Cees Nooteboom once wrote: ``Conversations consist for the most part of things one does not say'' \citep{Nooteboom:1996uq}. }  Extending the framework to incorporate multimodal features is a natural direction for future work.

Our focus is on settings in which the researcher can sample respondents into the study and can exercise some degree of control over the interlocutor.  This includes experimental designs in which the interlocutor is an AI system or a trained human actor.  These settings allow the researcher to manipulate aspects of the conversational process while observing how interactions unfold.

\subsection{Assignment, Policy, Conversation and History}

\noindent We set up our framework using standard potential outcomes notation, adapted to settings in which the treatment unfolds as a sequential, interactive process with latent features of interest.  Consider a respondent $i$ who is assigned to a conversational condition $Z_i \in \mathcal{Z}$.  We use $Z_i$ to denote the object randomized by the researcher.  Depending on the design, this assignment may determine the opening message, the conversational policy governing the interlocutor's behavior (e.g., the content of an AI system prompt), or both.\footnote{This set up extends naturally into observational settings as well in which $Z_{i}$ is not under researcher control, though with additional assumptions.  We focus on researcher-controlled experimental settings throughout this paper given the complexity of inference even when researchers have significant control over their design.  }  Respondents may differ in underlying characteristics, that may or may not be observable, such as preferences, beliefs, or conversational tendencies $U_{i}$, which can influence both the course of the interaction and the outcome.

For clarity of exposition, begin with the case in which the interlocutor speaks first.\footnote{The analysis can be straightforwardly extended to settings in which the respondent speaks first or in which the order of turns is otherwise structured.  In such cases, the initial history, $H_{i0}$, defined below, would include the respondent's opening message, and the policy would map that history into the interlocutor's first response. In the subsequent section, the main implication is that the \emph{opening message} estimand is unlikely to meet the identification assumptions because it is instead a standard message estimand endogenous to respondent utterances.}  Let $t$ index the turns of the conversation, where each turn is a pair of an interlocutor (AI) message $A_{it}$ and a respondent message $R_{it}$.  Thus, $A_{i0}$ denotes the initial message produced by the interlocutor.  Let the realized conversation be
\[
C_i = (A_{i0}, R_{i1}, A_{i1}, R_{i2}, \ldots, A_{iT}).
\]

\noindent We define the history of the conversation, $H_{it}$, available before the interlocutor produces message $A_{it}$ as
\[
H_{it} = (A_{i0}, R_{i1}, A_{i1}, R_{i2}, \ldots, R_{it}).\footnote{In practice, the influence of past utterances may be functionally constrained in certain ways---e.g., recency relevant---rather than fully dependent on the entire history.}
\]

\noindent The conversational history $H_{it}$ is not merely a record of past messages.  It is the object to which both the interlocutor and the respondent \emph{respond}.  In particular, respondent utterances may shape the subsequent behavior of both participants, \emph{including} of the respondent themselves.  This follows from work on how our own behaviors can affect subsequent beliefs, attitudes, and behaviors through mechanisms such as self-signaling and cognitive consistency \citep{Festinger:1962ao,Bem:1972du,Benabou:2011ml}.  As a result, the conversational process is path-dependent, with each utterance causally contributing to the conditions under which future utterances are generated.\footnote{Figure~\ref{fig:conversation-dag-unpacked} makes this causal structure more explicit.}

We assume that the interlocutor's behavior is governed by a conversational \emph{policy}, $\pi$.  A policy is a sequence of decision rules for the interlocutor\footnote{This definition does not assume that the rule is necessarily knowable, only that it exists.  In principle, the dependence of messages on conversational history could be represented through a lower-dimensional state variable, as in Markovian models.  However, in conversational settings, even if we theoretically assumed that such a state variable existed---a strong assumption---the observation of that state variable is even more challenging.  Because of that, they may be difficult to meaningfully specify, so we retain the full history for generality.  Doing so allows us to implicitly show what such a state variable would need to collapse and contain.}  The policy can capture a number of components of how an interlocutor is programmed to behave include, but not limited to, a generalized governing prompt, pre-determined responses at particular conversational moments, or components under the hood of the model, so to speak like temperature or how the model was fine-tuned.
\[
\pi = (\pi_0, \pi_1, \ldots, \pi_T),
\]
where each $\pi_t$ maps the conversational history $H_{it}$ into a message or a distribution over messages.  Under a deterministic policy,
\[
A_{it} = \pi_t(H_{it}),
\]
while under a stochastic policy,
\[
A_{it} \sim \pi_t(\cdot \mid H_{it}).
\]

\noindent Researcher assignment may determine the conversational policy.  We write $\pi_z$ to denote the policy induced by assignment condition $Z_i = z$.  This formulation accommodates a range of designs.  In one design, assignment randomizes only the initial message as a function of $Z_i$, while holding the policy fixed across respondents, so that $A_{i0} = g_0(Z_i)$ and $\pi_z = \pi$ for all $z \in \mathcal{Z}$.  In another, assignment randomizes the policy itself, so that $\pi_z$ varies with $z$, and the initial message is generated as the first output of that policy, $A_{i0} \sim \pi_0(\cdot \mid H_{i0})$.  More generally, assignment may determine both the opening message and the policy that governs the subsequent interaction.

The realized conversation $C_i$ is therefore generated by the opening message, the conversational policy, and the respondent's own behavior over the course of the interaction.  Even when $Z_i$ is randomized (as the next section shows), the conversation that unfolds is not fixed by assignment alone.  Respondents may produce different messages to the same opening message, and those responses shape the histories to which the interlocutor's policy subsequently responds.

\subsection{Potential Outcomes}

Finally, let $Y_i$ denote an outcome of interest measured after the conversation. Outcomes may be attitudes or behavior depending on the precise research question.\footnote{The assumption that outcomes occur after the entirety of the conversation simplifies our framework.  Obviously, researchers could be interested in outcomes that occur as a conversation progresses.  Within this framework, one could capture such outcomes by simply defining a particular history as the fully realized conversation.  However, if such outcomes are dynamic and constitute a sequence of co-occurring behaviors or attitudes, the framework would need to be adapted.  Additionally, theorizing about conversation features themselves as outcomes would involve an extension of our framework that we currently leave aside.  However, for examples, see \cite{Zhang:2018ik} and \cite{Tuan:2022hn}.}

Importantly, the way in which the conversation matters for outcomes may not necessarily be in terms of the full conversational path $C_i$---as a sequence of ordered collections of tokens, their timing, and who produced them---but instead due to \emph{latent} features, or \emph{representations} of that conversation, such as its level of incivility, empathy, or persuasiveness \citep{Egami:2022ax,Fong:2023um}.  As the literature notes, the conversation (even were it to be randomized) generates outcomes through a combination of features that the researcher measures directly or models, and other features of the conversation that remain unmodeled, whether because they are latent or simply not included in the analysis.  When the latent feature is of interest, but the estimand is constructed in terms of the conversation that generates it, confounding may occur as a consequence of unmeasured features of the conversation that are unaccounted for.

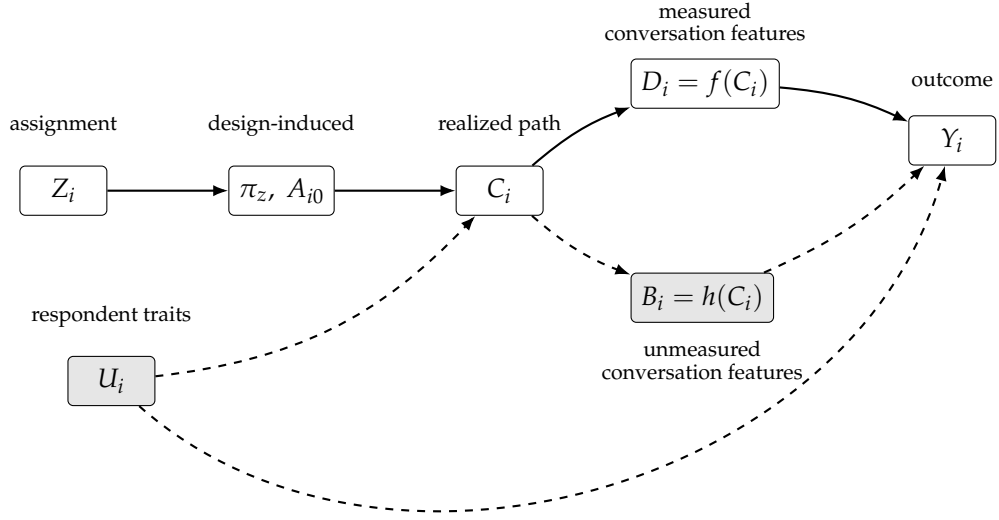
\begin{figure}[t]
	\centering
			\begin{tikzpicture}[
			    node distance=1.25cm and 1.6cm,
			    every node/.style={font=\small},
			    obs/.style={draw, rounded corners=2pt, minimum height=.65cm, minimum width=1.15cm, align=center, fill=white},
			    latent/.style={draw, rounded corners=2pt, minimum height=.65cm, minimum width=1.15cm, align=center, fill=gray!20},
			    arrow/.style={-{Latex[length=2mm]}, thick},
			    dashedarrow/.style={-{Latex[length=2mm]}, thick, dashed}
			]

			\node[obs] (Z) {$Z_i$};
			\node[obs, right=of Z] (policy) {$\pi_z,\ A_{i0}$};
			\node[obs, right=of policy] (C) {$C_i$};

			\node[obs, above right=.75cm and 1.15cm of C] (D) {$D_i=f(C_i)$};
			\node[latent, below right=.75cm and 1.15cm of C] (B) {$B_i=h(C_i)$};

			\node[obs, right=1.7cm of D, yshift=-.75cm] (Y) {$Y_i$};
			\node[latent, below=1.85cm of policy, xshift=-2.25cm] (U) {$U_i$};

			\draw[arrow] (Z) -- (policy);
			\draw[arrow] (policy) -- (C);
			\draw[arrow] (C) to[bend left=10] (D);
			\draw[dashedarrow] (C) to[bend right=10] (B);
			\draw[arrow] (D) to[bend left=10] (Y);
			\draw[dashedarrow] (B) to[bend right=10] (Y);
			\draw[dashedarrow] (U) to[bend right=20] (C);
			\draw[dashedarrow] (U) to[bend right=58] (Y);

			\node[above=.25cm of Z] {\scriptsize assignment};
			\node[above=.25cm of policy] {\scriptsize design-induced};
			\node[above=.25cm of C] {\scriptsize realized path};

			\node[above=.40cm of D] {\scriptsize measured};
			\node[above=.10cm of D] {\scriptsize conversation features};
			\node[below=.10cm of B] {\scriptsize unmeasured};
			\node[below=.40cm of B] {\scriptsize conversation features};
			\node[below=.95cm of B] {\scriptsize  };

			\node[above=.25cm of U] {\scriptsize respondent traits};
			\node[above=.25cm of Y] {\scriptsize outcome};

			\end{tikzpicture}
			\vspace{-.45in}
	\caption{Reduced form Directed Acyclic Graph (DAG) of a conversation. Shaded nodes are unobservable. Dashed arrows indicate relationships involving unobserved components. Observed individual traits $U_{i}$ are excluded for parsimony. Elements of this DAG are unpacked in subsequent figures. }
	\label{fig:conversation-dag}
\end{figure}

To capture this distinction between measurable and unmeasurable features of text (some of which may be latent), let $f : \mathcal{C} \rightarrow \mathcal{D}$ map a conversation into a measured feature of interest and let $h : \mathcal{C} \rightarrow \mathcal{B}$ denote unmeasured conversational features.  Define $D_i = f(C_i)$ and $B_i = h(C_i)$.\footnote{In some applications, features may be defined with respect to the history of the conversation up to a given point, so that $D_{it} = f(H_{it})$ captures aspects of the interaction as it unfolds over time.}

We assume that the outcome of interest depends on the conversation through these features, so that
\[
Y_i(C_i) = Y_i(D_i, B_i).
\]

\noindent This representation emphasizes that the realized conversation $C_i$ is the underlying object through which assignment affects outcomes, while the things researchers often care about may be abstract, lower dimensional features derived from that process.  Importantly, these features are not only properties of the completed conversation.  They also arise within the interaction itself.  At each point in the sequence, both observed and unobserved features of prior messages may shape subsequent responses.  In this sense, conversational features both summarize the realized path \emph{and} dynamically help to generate it.

Because $D_i$ and $B_i$ are jointly determined by the same underlying conversational path, variation in $D_i$ induced by changes in $C_i$ will in most cases be accompanied by variation in $B_i$. As a result, even in experimental settings, the measured feature $D_i$ may be systematically related to unmeasured features $B_i$ that also affect outcomes.  This interdependence is central to the inferential challenges that we discuss below.

Potential outcomes may therefore be indexed by different objects derived from this process.  One may define potential outcomes with respect to assignment, $Y_i(z)$; with respect to an opening message, $Y_i(a_0)$; with respect to a conversational policy, $Y_i(\pi)$; with respect to a message within the history $Y_i(H_{t},a_{t})$; or with respect to conversational features, $Y_i(d)$, among others.  These objects correspond to different theoretical questions and, in general, will not coincide.

Figure~\ref{fig:conversation-dag} summarizes this setup in a reduced form Directed Acyclic Graph (DAG) representation.  It omits individual messages within the conversation for clarity.  (Refer to subsequent DAGs, Figures~\ref{fig:conversation-dag-unpacked} and~\ref{fig:history-latent-dag} in this paper that bring individual messages in to develop the causal logic.)  Assignment $Z_i$ determines features of the design, such as the opening message $A_{i0}$ and/or the policy $\pi_z$, which help generate the realized conversation $C_i$.  The conversation then gives rise to measured and unmeasured features $(D_i, B_i)$, both of which may affect outcomes.

As illustrated by the figure, we also explicitly allow for both the realized conversation ($C_i$) and the outcome ($Y_i$) to be influenced by features of the respondent $U_i$. This causal arrow from $U_i$ to $C_i$ (and thus the downstream quantities $D_i$ and $B_i$) is the central inferential challenge: even when assignment is randomized, the conversational features of interest causally depend on respondent features.

\subsection{Context}

\noindent In this framework, it is useful to distinguish between different notions of \emph{context}, which other scholars of conversation and causality treat as critically important \citep{Zhang:2020qp,Zhang:2021zq}.  This framework decomposes context into three distinct parts.

First, the conversational history $H_{it}$ captures the sequence of prior utterances within the conversation.  This is often what is colloquially referred to as the context of a given message, but we treat it as a specifically efined object.  Second, respondent-specific characteristics $U_i$ influence both how individuals engage in the conversation and how they respond to it, and therefore play a central role in the identification challenges we describe.  The respondent is therefore part of the context in the sense of jointly producing the conversation and outcome.

Finally, there may exist broader situational context $X_i$---such as the environment or institutional setting---in which the conversation takes place.  For instance, the same conversation, between the same actors, that occurs in a dive bar on a Tuesday night, or on a Swiss mountaintop during a snowstorm will likely interpretively \emph{mean} different things \citep{Wittgenstein:1953qp}. We would then expect those meanings to cause different outcomes.  In many applications, particularly in AI-mediated experimental settings, we can reasonably treat this broader context as fixed---assuming that actors are, say, interacting in a homogenous graphical user interface in a laboratory---focusing instead on the interactional and individual sources of variation.\footnote{When we dive into estimands and their identifying assumptions, the framework assumption that we make to set aside $X_i$ can be thought of as part of the consistency assumption.}

The next section illustrates this set up with a single concrete example.

\subsection{An Anchoring Example: A Civil AI Chatbot}

\noindent Throughout the remainder of the paper, we will return to the example of the researcher is interested in civil political conversation and its relationship to attitudes towards democratic institutional integrity.\footnote{Along the lines of \cite{Allamong:2025ffq}.}  Recall that in this example, respondents are recruited into a study and randomly assigned to interact with an AI chatbot about a political topic.  Some respondents are assigned to a chatbot configured to engage in a civil and respectful manner, while others are assigned to a chatbot configured to engage in a more confrontational or uncivil manner.  After the conversation, respondents are asked their attitudes towards democratic institutions.

At first glance, the causal question appears straightforward---does a civil conversation increase perceptions that democracy has integrity? However, several distinct causal objects are potentially implicated. The researcher randomizes assignment to a conversational condition ($Z_i$) that might be casually referred to as the civil or uncivil condition. That assignment involves a configuration---an AI policy that governs how the interlocutor behaves throughout the interaction ($\pi_z$). Such a policy might include a prompt instructing the chatbot to remain civil, respectful, and non-confrontational; a specific opening message ($A_{i0}$) that initiates the discussion; a set of rules governing how the chatbot responds to disagreement, uncertainty, or hostility; constraints on the length or structure of responses; instructions about whether to ask follow-up questions or provide factual information; and even model-level settings such as temperature or retrieval procedures. More generally, the policy specifies not merely what is said, but how messages are generated as the conversation unfolds.

The policy then generates a sequence of interlocutor messages ($A_{it}$) in response to past respondent and interlocutor messages ($R_{it}$), producing a realized conversation ($C_i$). The theoretical object of interest, however, may not be the policy itself. A researcher may instead care about civility, represented as a conversational feature ($D_i=f(C_i)$) or as a feature of particular messages ($D_i=f(A_i)$). In that case, the policy serves as a mechanism for \emph{inducing} variation in civility, while civility itself remains the substantive causal object of interest. Distinguishing between these objects is central to the framework developed in this paper.

These distinctions matter because each corresponds to a different counterfactual question (see Table~\ref{tab:conversational_estimands}).  One might ask whether assignment to a civil chatbot increases participation; whether a conversational policy that encourages civility increases participation; whether a particular civil message increases participatory inclination; whether cumulative exposure to civil interaction increases participation, or whether one realized conversation would have produced different outcomes than another.  Although these questions are related, they are not identical.  The framework developed below clarifies how each corresponds to a distinct causal estimand and what assumptions are required to interpret it.

\begin{table}[t]
			\centering
			\scriptsize

			\begin{tabular}{p{3cm} p{4cm} p{5.5cm} p{3cm}}
			\toprule
			\textbf{Estimand} & \textbf{Form} & \textbf{Substantive question} & \textbf{Intervention} \\
			\midrule

			Assignment effect &
			$\tau_Z = \E[Y_i(z) - Y_i(z')]$ &
			What is the effect of being assigned to a conversational condition? &
			Assignment $Z_i$ \\

			\noalign{\smallskip}\noalign{\smallskip}

			Opening message effect &
			$\tau_{A_0} = \E[Y_i(a_0) - Y_i(a_0')]$ &
			How does the initial framing of a conversation affect outcomes? &
			Opening message $A_{i0}$ \\

			\noalign{\smallskip}\noalign{\smallskip}

			Policy effect &
			$\tau_\pi = \E[Y_i(\pi) - Y_i(\pi')]$ &
			What is the effect of deploying one conversational system rather than another? &
			Policy $\pi$ \\

			\noalign{\smallskip}\noalign{\smallskip}

			Message effect (conditional history) &
			$\tau_{A_t} = \E[Y_i(H_{it}, a_t) - Y_i(H_{it}, a_t')]$ &
			How would outcomes differ if a message changed, holding history fixed? &
			Message $A_{it}$, given $H_{it}$  \\

			\noalign{\smallskip}\noalign{\smallskip}

			Conversation effect &
			$\tau_C = \E[Y_i(c) - Y_i(c')]$ &
			What is the effect of experiencing one conversation rather than another? &
			Full path $C_i$ \\

			\noalign{\smallskip}\noalign{\smallskip}

			Feature effect &
			$\tau_D = \E[Y_i(d) - Y_i(d')]$ &
			What is the effect of a conversational feature (e.g., incivility)? &
			Feature $D_i = f(C_i)$ \\

			\noalign{\smallskip}\noalign{\smallskip}

			Dosage (exposure) effect &
			$\tau_E = \E[Y_i(e) - Y_i(e')]$ &
			What is the effect of cumulative exposure to a feature? &
			Exposure $E_i = g(D_{it})$ \\

			\noalign{\smallskip}\noalign{\smallskip}

			Representation-based CATE &
			$\tau^{CATE}_{A_t}(a,a' \mid S)$ &
			What is the message effect among similar conversational contexts? &
			Message $A_{it}$, given $S_{it}$ \\

			\noalign{\smallskip}\noalign{\smallskip}

			AMCE-style message effect &
			$\tau^{AMCE}_{A_t} = \E_S[\tau^{CATE}(a,a' \mid S)]$ &
			What is the average message effect across conversational contexts? &
			Message $A_{it}$ \\
\bottomrule
\end{tabular}
\caption{Causal estimands in conversational settings. Each estimand corresponds to a different intervention on the conversational process and answers a distinct substantive question.}
\label{tab:conversational_estimands}
\end{table}

The subsequent sections motivate and formalize these causal objects and estimands derived from this framework and illustrated by this example. We take each of these objects in turn.

\section{The effect of assignment to a general conversational condition}

\noindent A natural starting point to anchor a discussion of estimands is the effect of assignment to a conversational condition $Z_{i}$.  Substantively, this object captures how outcomes differ when respondents are assigned to different generalized conditions.  These conditions may be an initial conversational prompt, a particular interlocutor (person, model, so forth), or a particular conversational policy.  It is also often a bundle of these different components.  In our anchoring example, the assignment was a bundle of an initial message that was either civil and uncivil, as well as an AI policy that consisted of specific rules about how the AI should respond to respondent-generated messages.

Therefore, in practice, the assignment $Z_i$ is often a coarse object.  Researchers frequently interpret differences across assignment conditions as evidence about more specific objects---such as the effect of a particular message, tone, or style of interaction---\emph{even though} those objects are not directly manipulated by $Z_i$ alone.  This quantity is, however, closely aligned with experimental design and is often reported as the primary result in conversational experiments.  We discuss the assignment estimand as a kind of baseline to anchor later discussion of more precise estimands---it is what is identified under standard experimental designs, but it is often not the object of substantive interest.

Formally, let $Y_i(z)$ denote the potential outcome that would be observed for respondent $i$ under assignment $Z_i = z$.  The causal estimand of interest is the average treatment effect of assignment condition:
\[
\tau_Z(z,z') = \E[Y_i(z) - Y_i(z')].
\]

\noindent To interpret this quantity causally, we require the following standard assumptions, which we build on for later estimands.\footnote{We state these assumptions with respect to assignment here.  In what follows in the next sections, we apply them to different conversational objects---including opening messages, conversational policies, and so forth---with the understanding that the relevant intervention changes while the structure of the assumptions remains the same.}

\medskip\medskip

\noindent \emph{\textbf{Assumption 1 (Consistency and well-defined interventions).} For each $z \in \mathcal{Z}$, assignment to condition $z$ corresponds to a well-defined intervention, and the observed outcome satisfies $Y_i = Y_i(z)$ when $Z_i = z$.  This requires that assignment uniquely determines the relevant aspects of the intervention.}

\medskip

\noindent \emph{\textbf{Assumption 2 (No interference).} The potential outcome for respondent $i$ depends only on their own assignment, so that $Y_i(z)$ is unaffected by $Z_j$ for $j \neq i$.}

\medskip

\noindent \emph{\textbf{Assumption 3 (Ignorability of assignment).} Assignment is independent of potential outcomes:}
\[
\{Y_i(z): z \in \mathcal{Z}\} \perp Z_i.
\]

\medskip

\noindent \emph{\textbf{Assumption 4 (Positivity).} For all $z \in \mathcal{Z}$, $\Pr(Z_i = z) > 0$.}

\medskip\medskip

\noindent Under these assumptions, the assignment effect is identified by the difference in conditional expectations:
\[
\tau_Z(z,z') = \E[Y_i \mid Z_i = z] - \E[Y_i \mid Z_i = z'].
\]

\medskip

\noindent This result follows from standard arguments.  Importantly, the assignment estimand captures the effect of being assigned to a conversational condition. This is distinct from the effect of the conversation \emph{itself} or the features of conversation that the assignment is intended to induce. That is because those desired conversational features may not necessarily be successfully induced.  Indeed, the extend to which they are successfully induced may vary significantly as a function of respondent behavior---as in our example of the de-escalating interlocutor therapist or our escalating interlocutor internet troll. Thus, this estimand can be considered an intent to treat effect.

It is important to note that this is a substantial conceptual shift away from the theoretical object of interest (the nature of the conversation) in the name of identification.  Moreover, if one wishes to estimate an average treatment effect from such a design, one is limited to estimating a \emph{local} average treatment effect (LATE), one that is local to the population of participants who ``comply'' with the treatment. In our example of the therapist and internet troll, that means one is only able to estimate a LATE of incivility political attitudes for the type of people who, like the internet troll, play into the incivility of the AI, and not for the type of people who, like the therapist, might de-escalate an incivil conversation.

\section{The effect of the opening message}

\noindent A natural extension of the assignment estimand is the effect of the opening message itself.  Substantively, researchers are often interested not just in whether a respondent was exposed to a conversational condition, but in \emph{how that condition begins}.  In many settings, the opening message serves as the initial framing of the interaction: it may signal tone (civil or hostile)---as in our anchoring example---intent (persuasive or informational), or identity (in-group or out-group speaker).  These initial signals are theoretically interesting because of path dependence---they can shape not only immediate reactions but the \emph{trajectory} of the conversation that follows, albeit in non-deterministic ways.

This object is distinct from the assignment estimand when the mapping from assignment to opening message is not deterministic, or where assignment governs more than just the opening message (e.g., a system prompt that directs the AI to always be incivil).  For example, a researcher may assign respondents to an AI system configured to behave “uncivilly,” as in our anchoring example, but the specific message that initiates the conversation may vary across respondents due to stochastic generation or contextual variation.  In such cases, assignment identifies the effect of exposure to a class of conversational regimes bundled with specific messages, while the opening message captures a more precise and potentially more theoretically relevant object.

Recall that $A_{i0}$ denotes the opening interlocutor message.  When the opening message is directly determined or randomized by the researcher, we may define potential outcomes indexed by the message $Y_i(a_0)$, and the corresponding causal estimand,
\[
\tau_{A_0}(a_0,a_0') = \E[Y_i(a_0) - Y_i(a_0')].
\]

\noindent Under assumptions analogous to those in the assignment case but made with respect to opening message $A_{i0}$, this estimand is identified when opening messages are randomized across respondents.

Even if only the opening message is randomized, in many applications, the theoretical object of interest is not the exact message $a_0$, but a latent \emph{feature} of that message, such as its level of incivility, emotional tone, or group-signalling content.  As in the case of conversational features writ large, we may represent these as functions of the message.  Let $f_0 : \mathcal{A} \rightarrow \mathcal{D}_0$ denote a mapping from opening messages to a measured feature of interest and let $h_0 : \mathcal{A} \rightarrow \mathcal{B}_0$ denote unmeasured features.  Define $D_{i0} = f_0(A_{i0})$ and $B_{i0} = h_0(A_{i0})$.

We may then define potential outcomes indexed by the feature, $Y_i(d_0)$, and the corresponding estimand,
\[
\tau_{D_0}(d_0,d_0') = \E[Y_i(d_0) - Y_i(d_0')].
\]

\noindent Even when opening messages are randomly assigned, this estimand is not generally identified \citep{Egami:2022ax,Fong:2023um}.  The reason is not that randomization fails, but that the object of theoretical interest differs from what is randomized.  A single message bundles multiple outcome-relevant components.  When the researcher varies $A_{i0}$, they simultaneously vary both the feature of interest $D_{i0}$ and other aspects of the message captured by $B_{i0}$.  As a result, variation in $D_{i0}$ is not isolated from variation in $B_{i0}$, and differences in outcomes across values of $D_{i0}$ reflect the joint effect of both.

This distinction directly mirrors the problem of latent feature identification in textual treatments, generally.  The core requirement for identification is that variation in the feature of interest be decoupled from other aspects of the message that affect outcomes.  In particular, identification requires that either (i) the feature fully captures all outcome-relevant variation in the message, or (ii) variation in the feature is as-if randomized with respect to unmeasured components.

In practice, these assumptions are quite strong.  In our example, an initial message that differs in its civility, may also differ in other latent features that are related to the outcome, respondent's perception of democratic integrity---emotional intensity, expertise, implied speaker identity.  Unless these additional features are accounted for---either measured or controlled by design---the effect of incivility that is of researcher interest cannot be separated from the covariation.

This highlights an important difference between the effect of messages and the effect of features of messages.  While randomization of messages is sufficient to identify $\tau_{A_0}$, identifying $\tau_{D_0}$ requires additional assumptions or design features that isolate variation in the feature of interest.  This is not a mere technicality, but instead a deep component of what we, as researchers theorize about, and the interpretation we offer our estimates, even from randomized designs.  As we show in subsequent sections, these challenges become more pronounced once the conversation unfolds beyond the opening message.

\section{The effect of the conversational policy}

\noindent In conversations involving AI, researchers often have control over what computer science refers to as an AI \emph{policy}, and are interested in the effects of that policy.  A policy may correspond to a system prompt, a model configuration, a script, or any rule governing how messages are generated from conversational histories.  In conversations with AI, in particular, examples include instructing a model to have a particular tone---“empathetic,” “persuasive,” or “adversarial”---prompting discussion content or length, adjusting sampling parameters such as temperature, or fine-tuning a model on particular training data.\footnote{In human settings, a policy might correspond to a canvassing script, a deliberative protocol, or an interviewer's instructions. }  One may even program the AI to respond in specific ways at particular points in a conversation, or in response to certain types of respondent messages.  Crucially, in all of these cases, the policy specifies not just \emph{what is said}, but how responses are generated \emph{as the interaction unfolds}.

We can then consider potential outcomes indexed by policy, $Y_i(\pi)$, which denote the outcome that would be observed if the interaction were governed by policy $\pi$.  The causal estimand of interest is:
\[
\tau_{\pi}(\pi, \pi') = \E[Y_i(\pi) - Y_i(\pi')].
\]

\begin{figure}[t]
\centering
\begin{tikzpicture}[
    node distance=1.6cm and 1.8cm,
    every node/.style={font=\small},
    inter2/.style={draw, rounded corners=4pt, minimum height=.6cm, minimum width=1.2cm, align=center},
    resp2/.style={draw, minimum height=.6cm, minimum width=1.2cm, align=center},
    obs/.style={draw, rounded corners=2pt, minimum height=.7cm, minimum width=1.3cm, align=center},
    latent/.style={draw, rounded corners=2pt, minimum height=.7cm, minimum width=1.3cm, align=center, fill=gray!20},
		inter/.style={draw, rounded corners=2pt, minimum height=.7cm, minimum width=1.3cm, align=center, fill=blue!50!purple!60!gray!10!},
		resp/.style={draw, rounded corners=2pt, minimum height=.7cm, minimum width=1.3cm, align=center, fill=teal!80!gray!10},
    arrow/.style={-{Latex[length=2mm]}, thick},
    dashedarrow/.style={-{Latex[length=2mm]}, thick, dashed},
		aiarrow/.style={-{Latex[length=2mm]}, thick, draw=purplestrong!80!black},
		resparrow/.style={-{Latex[length=2mm]}, thick, draw=tealstrong!80!black},
]

\node[obs] (pi) {$\pi$};

\node[inter, right=of pi] (I0) {$A_{i0}$};
\node[resp, right=of I0] (R0) {$R_{i0}$};

\node[inter, right=of R0] (I1) {$A_{i1}$};
\node[resp, right=of I1] (R1) {$R_{i1}$};

\node[obs, below=3.3cm of R0, xshift=1.5cm] (C) {$C_i$};

\node[obs, right=2.5cm of C, yshift=.2cm] (D) {$D_i=f(C_i)$};
\node[latent, right=2.5cm of C, yshift=-1.4cm] (B) {$B_i=h(C_i)$};

\node[obs, right=of D, yshift=-0.7cm] (Y) {$Y_i$};

\node[latent, below=3.5cm of pi, xshift=1.8cm] (U) {$U_i$};

\draw[arrow] (pi) -- (I0);
\draw[arrow] (pi) to[bend right=25] (I1);

\draw[aiarrow] (I0) -- (R0);
\draw[resparrow] (R0) -- (I1);
\draw[aiarrow] (I1) -- (R1);

\draw[aiarrow] (I0) to[bend right=25] (C);
\draw[resparrow] (R1) to[bend left=25] (C);
\draw[aiarrow] (I1) to[bend right=5] (C);
\draw[resparrow] (R0) to[bend left=5] (C);

\draw[arrow] (C) to[bend left=5] (D);
\draw[arrow] (C) to[bend right=10] (B);

\draw[arrow] (D) to[bend left=10] (Y);
\draw[dashedarrow] (B) to[bend right=10] (Y);

\draw[dashedarrow] (U) to[bend right=15] (R0);
\draw[dashedarrow] (U) to[bend right=15] (R1);
\draw[dashedarrow] (U) to[bend right=55] (Y);
\draw[aiarrow] (I0) to[bend right=15] (I1);
\draw[aiarrow] (I0) to[bend right=25] (R1);
\draw[resparrow] (R0) to[bend right=15] (R1);

\node[above=0.6cm of I0, xshift=1.5cm] (t0) {$\overbrace{\hspace{4.7cm}}^{\text{turn } t=0}$};
\node[above=0.6cm of I1, xshift=1.5cm] (t1) {$\overbrace{\hspace{4.7cm}}^{\text{turn } t=1}$};

\node[above=.3cm of pi] {\scriptsize policy};
\node[above=.3cm of I0] {\scriptsize interlocutor};
\node[above=.02cm of I0] {\scriptsize message};
\node[above=.3cm of R0] {\scriptsize respondent};
\node[above=.03cm of R0] {\scriptsize message};

\node[above=.3cm of I1] {\scriptsize interlocutor};
\node[above=.02cm of I1] {\scriptsize message};
\node[above=.3cm of R1] {\scriptsize respondent};
\node[above=.03cm of R1] {\scriptsize message};

\node[above=.45cm of D] {\scriptsize measured };
\node[above=.15cm of D] {\scriptsize conversational features};
\node[below=.15cm of B] {\scriptsize unmeasured };
\node[below=.45cm of B] {\scriptsize conversational features};
\node[below=.85cm of B] {\scriptsize };

\node[above=.15cm of U] {\scriptsize respondent traits};
\node[above=.15cm of Y] {\scriptsize outcome};

\end{tikzpicture}
\vspace{-.35in}
\caption{Directed Acyclic Graph (DAG) that unpacks the conversation process with two turns. The policy governs interlocutor messages at each turn, while respondent behavior drives the evolution of the interaction. The treatment regime $Z_{i}$ is excluded only for expositional clarity.  The conversational path $C_i$ is included as a notational device that summarizes the sequence of interlocutor and respondent messages.  It is not its own seperate causal variable; rather, $C_i$ is a deterministic function of $(I_{i0}, R_{i0}, I_{i1}, R_{i1}, \ldots)$.  The figure uses $C_i$ to simplify the representation of downstream features $D_i$ and $B_i$, though all causal relationships could be written directly in terms of the underlying sequence of messages (see Figure~\ref{fig:history-latent-dag}).  In a fully expanded representation, features of prior messages ($D_{it}$ and $B_{it}$) would enter into the generation of subsequent messages, though we suppress this structure in the figure for clarity.  Observed individual traits $U_{i}$ are also excluded for parsimony.  Dashed arrows indicate relationships involving unobserved components.}
\label{fig:conversation-dag-unpacked}
\end{figure}
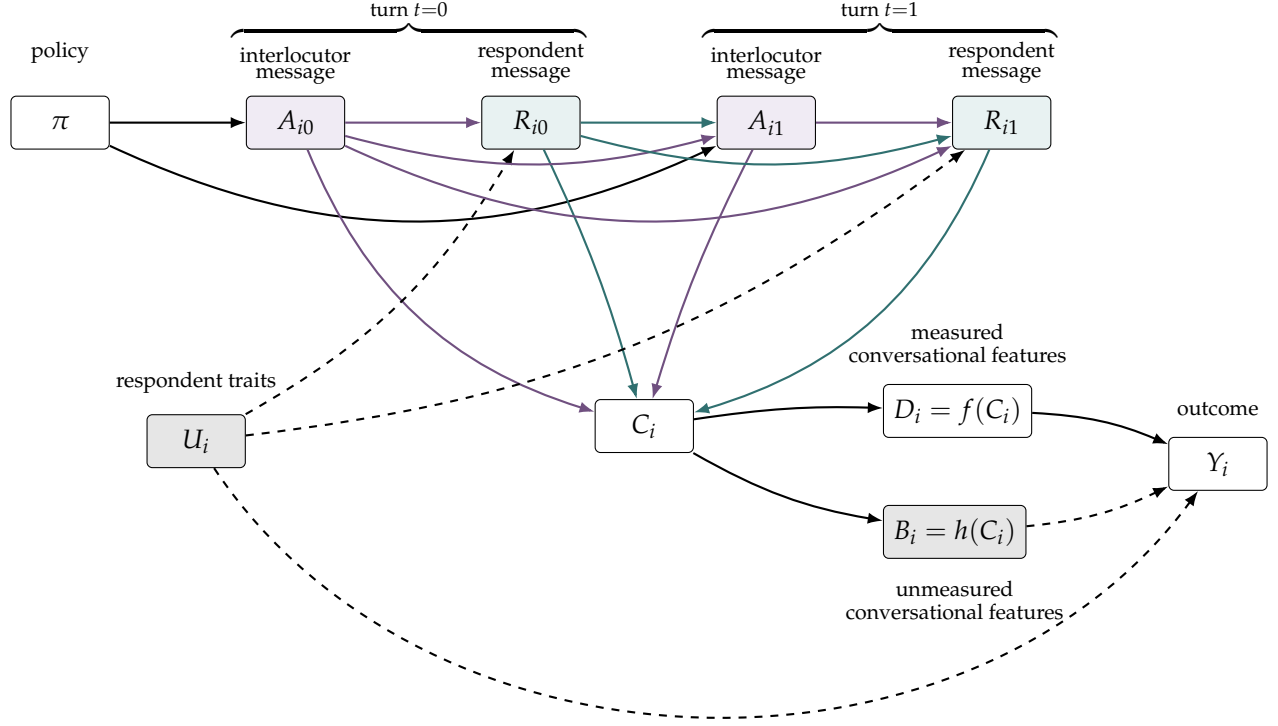

\noindent In the simplest case, a policy is effectively static: it determines a fixed opening message or a script that unfolds without regard to respondent behavior. In such settings, the policy collapses to a standard treatment and the policy effect is conceptually equivalent to the effect of assignment. More generally, however, policies are dynamic (Figure~\ref{fig:conversation-dag-unpacked} and~\ref{fig:history-latent-dag}). They specify how the interlocutor responds at each point in the interaction as a function of the evolving conversational history. Thus, a policy governs not a single message but an entire conversational process.

When policies are randomly assigned, the total effect $\tau_\pi$ is identified under Assumptions 1--4, interpreted with respect to assignment to policies, together with an additional condition:

\medskip\medskip

\noindent \emph{\textbf{Assumption 5 (Stable and independent policy implementation).} The policy assigned to respondent $i$ determines the conversational process they experience and is not affected by other respondents' interactions or prior conversations.}

\medskip\medskip

\noindent Assumption 5 extends the usual consistency requirement to conversational settings. It requires not only that the policy be well-defined, but that it be implemented consistently across respondents. In static experimental settings this may be plausible. In conversational settings it is more delicate. In human-mediated interactions, interlocutors may learn, adapt, or change their behavior across conversations (see Section~\ref{sec:humans}). In AI-mediated settings, similar issues arise when systems are updated over time, incorporate conversational memory, or rely on shared retrieval mechanisms. In our anchoring example, our AI model might learn that respondents who express skepticism about elections tend to respond favorably to particular arguments and then deploy those arguments in subsequent conversations. As a result, the conversational policy experienced by one respondent may depend on information learned from previous respondents. In such cases, the policy governing one interaction depends on earlier interactions, implying that the estimand no longer corresponds to the effect of any notion of a fixed conversational regime.\footnote{The way in which it depends may not be transparent in a way that would allow the dependent nature to be modelled.  }

A second feature of conversational policies is that they are often stochastic. The same conversational history $H_{it}$ can generate different messages under repeated realizations of the same policy. Consequently, a policy induces not a single conversational path but a \emph{distribution over possible conversational paths}. Accordingly, the potential outcome $Y_i(\pi)$ should therefore be understood as the outcome under a regime that generates a distribution of interactions rather than a single fixed conversation.

This perspective clarifies both the appeal and the limitation of the policy estimand. On the one hand, $\tau_\pi$ corresponds directly to a practically relevant question: what happens if one conversational system is deployed rather than another? On the other hand, because a policy induces many possible conversational trajectories, the policy effect \emph{averages} over interactions that may differ substantially in their attributes---e.g., content, tone, length, engagement. Thus, even when identified, the policy effect combines many aspects of the interaction into a single quantity.\footnote{One response to this aggregation is to move closer to local conversational interventions.  Section~\ref{sec:message-effect} develops conditional message-level estimands, which allow effects to vary across conversational contexts.  These estimands help unpack some of the heterogeneity that is averaged over by the policy effect, though they correspond to different causal objects and require stronger assumptions for identification.  More generally, characterizing the distributions of trajectories induced by different policies may itself constitute a distinct object of inquiry.}

Relatedly, policies may also bundle multiple conversational features. For example, in our anchoring example, a policy intended to encourage civil policy discussion may simultaneously increase perceived expertise of the interlocutor, emotional intensity of the message content, may increase the willingness of the respondent to engage, and may consequently generate a longer conversation.   But the exact same policy condition may generate a brief exchange for another respondent because the policy stochastically generates a different degree of civility in its messages that in turn have different emotional intensity and are engaged with differently by the respondent.  A positive policy effect therefore need not imply that civility itself is the operative mechanism, nor that respondents experienced similar conversations despite being assigned to the same policy.

Generally speaking then, policy effects are often difficult to interpret as the effect of any single conversational attribute. In this respect, policies resemble messages that we discuss in the next section in that both may bundle multiple theoretically relevant components into a single intervention.  Importantly, this is primarily a question of \emph{interpretation} rather than identification. The average policy effect may be straightforwardly identified under random assignment without requiring knowledge of the full distribution of messages induced by $\pi$. Yet there are ample reasons to think that that distribution still matters. It determines which conversational paths are plausibly occur, and therefore will be observed.  Two implementatins of the same policy may therefore produce similar average outcomes while generating very different conversational experiences.\footnote{This concern is reflected in a growing literature on the reproducibility of large language models, which shows that systems with identical prompts or configurations can generate different outputs across runs, or over time \citep{Liang:2022ik,Chen:2023eq,Bommsani:2023ey}.}

In sum, these considerations reinforce a central point.  Policy effects are often the most straightforward to identify, but they are also the most aggregated.  They describe what happens under a conversational regime, but not how that effect is produced.  Answering those questions requires moving from policies to the realized conversational process, which we consider next.

\section{The effect of a message within a conversation}\label{sec:message-effect}

\noindent A distinct causal object arises when the researcher moves from considering broad conversational regimes to being interested in specific message realizations---and features of them---within a conversation.  Rather than asking what happens when a policy is assigned, one may ask how outcomes would differ if a particular message within the interaction had been different.

At the message level, however, there are two conceptually distinct comparisons that can be made.  The first varies the message while holding the conversational history fixed.  The second holds the message fixed while allowing the history to vary.  Only the former corresponds to a well-defined causal intervention on the message itself.  The latter reflects variation in both message and context, and therefore conflates message effects with differences in conversational histories.

We focus on the first comparison, which isolates the effect of a message conditional on the history in which it occurs.  Consider potential outcomes indexed by the history and the message, $Y_i(H_{it}, a_t)$, which represent the outcome that counterfactually would be observed if the conversation evolved up to $H_{it}$ and the interlocutor then produced message $a_t$.\footnote{In practice, we think of this in a conditional structure where the effect of $a_t$ is conditional on history, but one could also be interested in history and message with different functional relationships.}  As noted in previous sections, the theoretical object of interest is often not the message itself, but features of that message.  We introduce these formally below, but note here that such features are functions of both the message and the conversational context in which it appears.  The corresponding causal estimand compares alternative continuations of the same conversational history:
\[
\tau_{A_t}(a_t, a_t' \mid H_{it}) = \E[Y_i(H_{it}, a_t) - Y_i(H_{it}, a_t')].
\]

\noindent This estimand has a natural interpretation in terms of conversational paths that branch.  Holding the prior history fixed, different messages at time $t$ generate different future conversational trajectories.

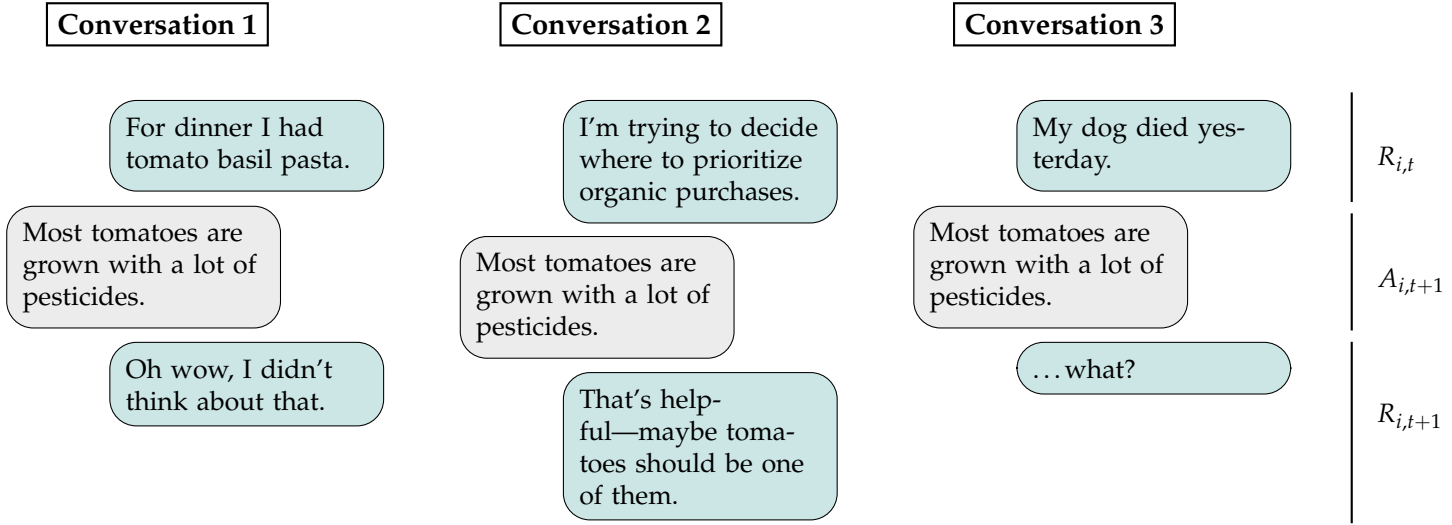
\begin{figure}[t]
			\centering
			\begin{tikzpicture}[
			    every node/.style={font=\small},
			    human/.style={
			        draw,
			        rounded corners=10pt,
			        fill=teal!20!white,
			        align=left,
			        inner sep=6pt,
			        text width=3.2cm
			    },
			    ai/.style={
			        draw,
			        rounded corners=10pt,
			        fill=gray!15,
			        align=left,
			        inner sep=6pt,
			        text width=3.2cm
			    },
			    title/.style={font=\bfseries}
			]

			\node[
			    draw,
			    rectangle,
			    thick,
			    inner sep=4pt,
			    font=\bfseries
			] (t1) at (0,0) {Conversation 1};
			\node[
			    draw,
			    rectangle,
			    thick,
			    inner sep=4pt,
			    font=\bfseries
			] (t2) at (6,0) {Conversation 2};
			\node[
			    draw,
			    rectangle,
			    thick,
			    inner sep=4pt,
			    font=\bfseries
			] (t3) at (12,0) {Conversation 3};

			\node[human, anchor=north east] at (3, -1)
			{For dinner I had tomato basil pasta.};

			\node[ai, anchor=north west] at (-2, -2.4)
			{Most tomatoes are grown with a lot of pesticides.};

			\node[human, anchor=north east] at (3, -4.2)
			{Oh wow, I didn’t think about that.};

			\node[human, anchor=north east] at (9, -1)
			{I’m trying to decide where to prioritize organic purchases.};

			\node[ai, anchor=north west] at (4, -2.8)
			{Most tomatoes are grown with a lot of pesticides.};

			\node[human, anchor=north east] at (9, -4.6)
			{That’s helpful—maybe tomatoes should be one of them.};

			\node[human, anchor=north east] at (15, -1)
			{My dog died yesterday.};

			\node[ai, anchor=north west] at (10, -2.4)
			{Most tomatoes are grown with a lot of pesticides.};

			\node[human, anchor=north east] at (15, -4.2)
			{…what?};

			\draw[thick] (15.8, -0.9) -- (15.8, -2.35);
			\draw[thick] (15.8, -2.5) -- (15.8, -4.05);
			\draw[thick] (15.8, -4.2) -- (15.8, -6.6);

			\node[anchor=west] at (16.0, -1.8) {$R_{i,t}$};
			\node[anchor=west] at (16.0, -3.4) {$A_{i,t+1}$};
			\node[anchor=west] at (16.0, -5.2) {$R_{i,t+1}$};

			\end{tikzpicture}
			\caption{The same AI message can have very different meanings depending on conversational context.  Although the textual content is identical, its interpretation---and therefore its causal effect---depends on the preceding interaction.  When we think of design, we would vary the AI message ($A_{i,t+1}$), but \emph{within} a message-level treatment arm we can still encounter this interpretive variation. }
			\label{fig:conversation-context}
\end{figure}

A key feature of this setup is that messages are not meaningful in isolation.  Even though this estimand concerns the effect of different messages, one can illustrate this conditionally generated message \emph{meaning} by considering how the same message text may carry very different meanings depending on the conversational context in which it appears.  For example, the same AI utterance may be interpreted as helpful, irrelevant, or even offensive depending on the respondent’s prior message or message history (Figure~\ref{fig:conversation-context}). Certain message features may be differentially perceived at different points within a conversation \citep{Li:2020oa}. This demonstrates that the causal effects of messages are inherently \emph{context-dependent}.  The relevant context is not background noise, but is itself generated within the conversation and shapes the latent features of a message that are often of theoretical interest.

Identifying $\tau_{A_t}$ therefore requires assumptions about how histories and messages relate to potential outcomes.  In particular, the central condition is a form of sequential ignorability.

\medskip\medskip

\noindent \emph{\textbf{Assumption 6 (Sequential history ignorability).} For all $t$ and all $a_t$, $A_{it} \perp Y_i(H_{it}, a_t) \mid H_{it}$.}

\medskip\medskip

\noindent Sequential ignorability requires that, conditional on the conversational history, the next message is as-if randomly assigned. That is, once we account for everything relevant about the interaction up to that point, there are no remaining factors that jointly influence both the next message and the eventual outcome. In principle, this assumption could hold if the observed history were a sufficient summary of the conversational state. In practice, however, the histories available to the researcher are jointly generated by the respondent and the interlocutor and may fail to capture latent respondent characteristics---such as evolving beliefs, engagement, or conversational style---that influence both subsequent messages and eventual outcomes (Figure~\ref{fig:conversation-dag-unpacked}). As a result, conditioning on the observed conversational history may not remove all confounding (see Proposition~1 later in this section).

\medskip\medskip

\noindent \emph{\textbf{Assumption 7 (Consistency for message interventions).} If $A_{it} = a_t$, then $Y_i = Y_i(H_{it}, a_t)$.}

\medskip\medskip

\noindent Consistency requires that the potential outcome indexed by $(H_{it}, a_t)$ corresponds to a well-defined intervention.  When the observed message equals $a_t$, the observed outcome must coincide with the corresponding potential outcome.  This implicitly assumes that the effect of a message is stable for a given $(H_{it}, a_t)$ pair.  In conversational settings, this assumption is nontrivial because the object of theoretical interest is often the meaning of the message rather than the text alone.  Meaning is context-dependent and arises from the interaction between the message and the conversational history.  While our framework captures this context through $H_{it}$ (or through features that we introduce below), the relevant aspects of that context may not be fully observable or measurable, and thus may not be fully accounted for in practice.

\medskip\medskip

\noindent \emph{\textbf{Assumption 8 (Positivity over histories).} For all relevant histories $H_{it}$ and messages $a_t$, $\Pr(A_{it} = a_t \mid H_{it}) > 0$.}

\medskip\medskip

\noindent Positivity requires that, for any history of interest, there is sufficient variation in the messages that follow.  If certain messages are never (or almost never) observed following particular conversational trajectories, then their effects cannot be learned from the data.

A further and related complication arises from the high-dimensional nature of conversational histories, especially if we think of them simply as \emph{text} \citep{Li:2020oa,Egami:2022ax}.  Even if positivity holds in principle, the space of possible histories may be so large that each realized history is effectively unique in the data.  In such settings, standard empirical notions of conditioning on $H_{it}$ provides little meaningful leverage, as there are no comparable observations under alternative messages.  This sparsity limits the feasibility of estimating message-level effects even when the formal assumptions are satisfied.

Let's consider these assumptions concretely in the context of our anchoring example.  If we wish to know the effect of a particular civil or uncivil AI message \emph{within} a conversation, the difficulty is that, while assignment to the civil AI chatbot condition may be randomized, the specific message that appears later in the conversation generally is not. Consequently, respondents who have certain (potential) outcomes in terms fo their perception of of democratic institutions may also be those who are more engaged, more receptive, or more conciliatory over the course of the interaction. Those tendencies shape what \emph{they} say, \emph{which in turn} shapes how AI interlocutor responds. As a result, an uncivil message observed later in a conversation may be endogenous to respondent characteristics that also affect the eventual outcome.

Second, in terms of consistency, imagine that two respondents receive the identical AI message: ``I understand your concern about democratic institutions.'' In one conversation, the message follows a long and respectful exchange and is interpreted as sincere validation. In another, it follows a series of dismissive responses and is interpreted as patronizing or sarcastic. The issue here is different from the one above. Even if we somehow solved the endogeneity problem (e.g., see Section~\ref{subsec:messagerandomization}) and could compare respondents receiving the same message, the treatment itself may not be stable in terms of it's meaning and interpretation for the respondent in that conversation.

Finally, with regard to the positivity assumption, imagine a respondent who has repeatedly expressed distrust in electoral institutions while the AI has consistently responded in a civil and empathetic manner. If an uncivil response is effectively never observed following such a history, then there is no empirical basis for learning what respondent outcome counterfactually \emph{would have} occurred had the conversation instead taken a different turn. More generally, conversational histories may be regarded as so specific that each realized interaction is effectively unique. In that case, even if the previous assumptions held conceptually, there may be no meaningful comparisons empirically available from which to estimate the effect of the message of interest.

Taken together, assumptions 6-8 define the conditions under which the message-level estimand $\tau_{A_t}(a_t, a_t' \mid H_{it})$ is identified.  They imply that identifying message effects requires histories that are sufficiently comparable, variation in messages conditional on those histories, and a well-defined interpretation of what it means to intervene on the message itself.  As our anchoring example illustrates, however, these conditions are unlikely to hold in practice.  The following propositions formalize two crucial sources of non-identification.

\medskip\medskip

\noindent \emph{\textbf{Proposition 1 (Selection bias from endogenous histories).}
The difference in observed outcomes between $A_{it} = a_t$ and $A_{it} = a_t'$ can be decomposed into the causal effect of the message conditional on history and a bias term arising from differences in conversational histories across message conditions.}

\medskip\medskip

\noindent The source of bias can be seen by comparing the estimand to what is observed in the data.  The causal effect of interest holds $H_{it}$ fixed, comparing $Y_i(H_{it}, a_t)$ and $Y_i(H_{it}, a_t')$.  In the observed data, however, messages arise from histories that differ across individuals.  As a result, comparisons of outcomes across $A_{it} = a_t$ and $A_{it} = a_t'$ implicitly compare
\[
Y_i(H_{it}, a_t) \quad \text{and} \quad Y_i(H_{it}', a_t'),
\]
where $H_{it} \neq H_{it}'$.  Because histories depend on respondent characteristics $U_i$ that also affect outcomes, this induces a bias term reflecting differences in $U_i$ across message conditions (see Appendix~\ref{appendix_proof_prop1}).  This follows from the causal structure:
\[
U_i \rightarrow R_{it} \rightarrow H_{it} \rightarrow A_{it}
\quad \text{and} \quad
U_i \rightarrow Y_i,
\]
so that $A_{it}$ is not independent of potential outcomes conditional on observed histories.  The respondent is not a passive recipient of messages; their behavior shapes the history that determines subsequent treatment.

A closely related issue arises when the researcher is interested not in the message itself, but in features of the message, such as tone, sentiment, or framing.  As discussed above, we can represent these features as functions of the message and its context, for example $D_{it} = f(H_{it}, A_{it})$ for observed features and $B_{it} = h(H_{it}, A_{it})$ for unobserved components.  In this formulation, the message serves as a vehicle through which multiple features are jointly realized.

\medskip\medskip

\noindent \emph{\textbf{Proposition 2 (Feature bundling bias).}
Comparisons of outcomes across values of a conversational feature reflect both the effect of the feature of interest and the effect of other outcome-relevant aspects of the message that co-vary with it.}

\medskip\medskip

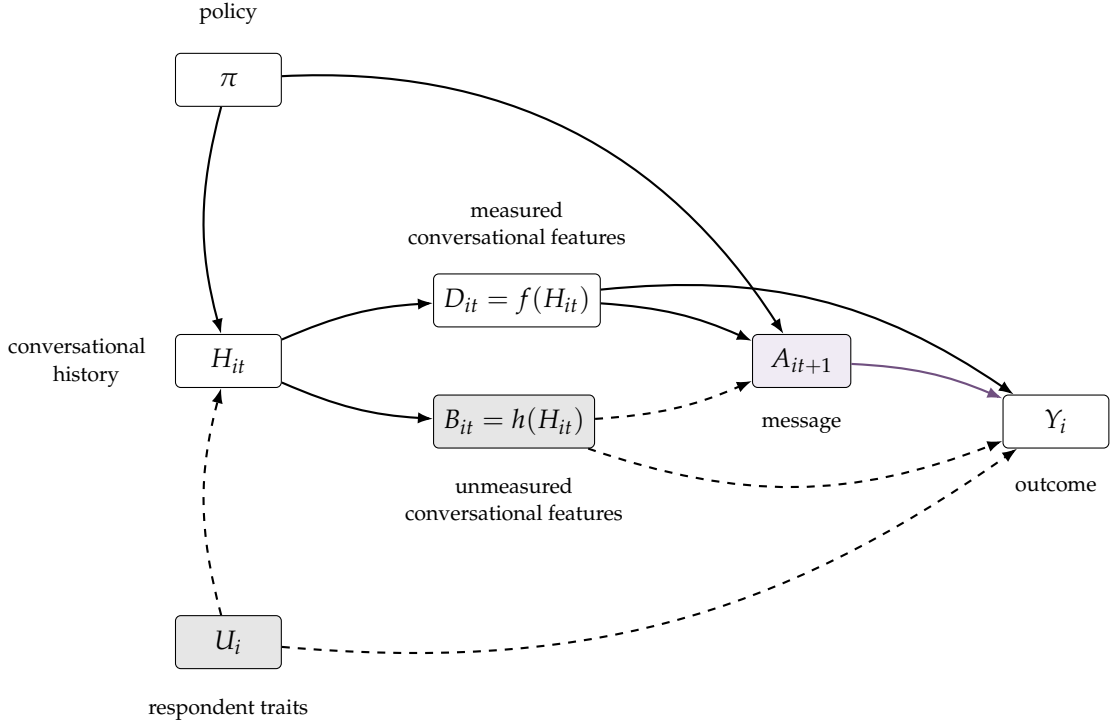
\begin{figure}[t]
\centering
\begin{tikzpicture}[
    node distance=1.6cm and 2.0cm,
    every node/.style={font=\small},
    obs/.style={draw, rounded corners=2pt, minimum height=.7cm, minimum width=1.4cm, align=center},
    latent/.style={draw, rounded corners=2pt, minimum height=.7cm, minimum width=1.4cm, align=center, fill=gray!20},
    arrow/.style={-{Latex[length=2mm]}, thick},
		inter/.style={draw, rounded corners=2pt, minimum height=.7cm, minimum width=1.3cm, align=center, fill=blue!50!purple!60!gray!10!},
		aiarrow/.style={-{Latex[length=2mm]}, thick, draw=purplestrong!80!black},
    dashedarrow/.style={-{Latex[length=2mm]}, thick, dashed}
]

\node[obs] (H) {$H_{it}$};
\node[obs, right=of H, yshift=0.8cm] (D) {$D_{it} = f(H_{it})$};
\node[latent, right=of H, yshift=-0.8cm] (B) {$B_{it} = h(H_{it})$};

\node[inter, right=of D, yshift=-0.8cm] (A) {$A_{it+1}$};

\node[obs, right=of A, yshift=-0.8cm] (Y) {$Y_i$};

\node[latent, below=3.0cm of H] (U) {$U_i$};
\node[obs, above=3.0cm of H] (pi) {$\pi$};

\draw[arrow] (H) to[bend left=10](D);
\draw[arrow] (H) to[bend right=10] (B);

\draw[arrow] (D) to[bend left=10] (A);
\draw[dashedarrow] (B) to[bend right=10] (A);

\draw[arrow] (D) to[bend left=20] (Y);
\draw[dashedarrow] (B) to[bend right=20] (Y);

\draw[aiarrow] (A) to[bend left=10] (Y);

\draw[dashedarrow] (U) to[bend left=15] (H);
\draw[dashedarrow] (U) to[bend right=20] (Y);

\draw[arrow] (pi) to[bend left=30] (A);
\draw[arrow] (pi) to[bend right=15] (H);

\node[left=.25cm of H,yshift=.2cm] {\scriptsize conversational };
\node[left=.6cm of H,yshift=-.2cm] {\scriptsize history };
\node[above=.6cm of D] {\scriptsize measured};
\node[above=.25cm of D] {\scriptsize conversational features};
\node[below=.25cm of B] {\scriptsize unmeasured};
\node[below=.6cm of B] {\scriptsize conversational features};
\node[below=0.7cm of U] {\scriptsize };
\node[below=.25cm of A] {\scriptsize message};
\node[below=.25cm of Y] {\scriptsize outcome};
\node[below=.25cm of U] {\scriptsize respondent traits};
\node[above=.25cm of pi] {\scriptsize policy};

\end{tikzpicture}
\caption{Directed Acyclic Graph (DAG) of conversational history and next interlocutor message.  The conversational history $H_{it}$ generates both observed features $D_{it}$ and unobserved features $B_{it}$, which jointly influence the next message $A_{it+1}$ and the outcome $Y_i$.  Respondent-specific traits $U_i$ shape both the evolution of the conversation and the outcome.  This representation emphasizes that ``history'' is not merely a textual record, but a structured object containing latent components that are relevant for both message generation and causal inference.  Measured and unmeasured that result from message $A_{it+1}$ are suppressed for parsimony, keeping the focus on the generation of the next interlocutor message.  Observed respondent traits and assignment $Z_{i}$ are suppressed for parisoony.  Dashed arrows indicated unobserved relationships.}
\label{fig:history-latent-dag}
\end{figure}

\noindent This second source of non-identification arises because messages bundle multiple outcome-relevant components, along the lines of \cite{Egami:2022ax,Fong:2023um}.  Even if messages were randomized, variation in any feature of interest would generally co-vary with other unmeasured aspects of the message (Figure~\ref{fig:history-latent-dag}).  As a result, differences in outcomes across feature values reflect the joint effect of measured and unmeasured components.  In order to identify feature-level effects one would require additional assumptions that isolate variation in the feature of interest from other aspects of the message.  In particular, this would require that all outcome-relevant components of the message be either observed or conditionally independent of the feature given the observed history.  In conversational settings, such conditions are unlikely to hold in practice.  We specify the issues more explicitly in Appendix~\ref{appendix_proof_prop2}.

These results together indicate a fundamental challenge.  Message-level estimands are often the most closely aligned with theoretical questions about persuasion, disagreement, or information processing.  However, they require conditioning on a history that is itself endogenously generated, and isolating variation in features that are bundled within messages.  These two sources of difficulty make message effects both theoretically appealing and empirically challenging to identify.

\section{The effect of the realized conversation}\label{sec:conversation-effect}

\noindent A final causal object is the effect of the realized conversation itself.  Here the researcher is not interested only in assignment, the opening message, the policy, or a single message within the interaction.  Rather, the object of interest is the full conversational path---what was said, in what order, by whom, and in response to what.  Additionally, the interest may be in features of that more structured conversation---the topic, the tone, the sentiment.

Substantively, this object corresponds to questions such as: what is the effect of having this conversation rather than that one?  What would have happened if the respondent had experienced a conversation that was more deliberative, more hostile, more informative, more empathetic, or more sustained?  These are often the questions researchers actually have in mind when they describe a conversation as the treatment.  They are also the questions that become most difficult once we take seriously that conversations are jointly produced rather than assigned.

Recall that $C_i$ denotes the realized conversational path.  We define potential outcomes indexed by the full conversation, where $Y_i(c)$ is the outcome that would be observed for respondent $i$ if they experienced conversational path $c$.  The corresponding estimand is:
\[
\tau_C(c,c') = \E[Y_i(c) - Y_i(c')].
\]

\noindent This estimand is conceptually straightforward but empirically extremely demanding.  It treats the conversation itself as the causal object.  Unlike the assignment estimand, which asks about being placed into a conversational condition, $\tau_C$ asks about the effect of the conversation that actually unfolds.  This distinction matters because two respondents assigned to the same condition may experience very different conversations.  A respondent may resist, redirect, escalate, disengage, joke, challenge, or disclose something unexpected. A respondent may have limited engagement and generate a short conversation; they may take over and the conversation may last a long time.  And so forth.  In doing so, they may walk the conversation quite far from where the researcher thought it would go.

This is not a small nuisance.  It is part of what \emph{makes} conversations conversations.  The realized path $C_i$ is generated by the policy, the interlocutor, the respondent, and the history that accumulates between them (Figures~\ref{fig:conversation-dag} and~\ref{fig:conversation-dag-unpacked}).  As a result, $C_i$ is post-assignment and endogenous.  Even when $Z_i$ is randomly assigned, the realized conversation is not generally as-if randomly assigned.

To identify $\tau_C(c,c')$ from observed variation in realized conversations, one needs to assume a distinct ignorability:

\medskip\medskip

\noindent \emph{\textbf{Assumption 9 (Conversation ignorability)}: $C_i \perp \{Y_i(c): c \in \mathcal{C}\}.$}

\medskip\medskip

\noindent These conditions are sufficient for identification of the conversation-level estimand $\tau_C(c,c')$.  If realized conversations were as-if randomly assigned, then differences in observed outcomes across conversational paths could be interpreted as causal effects of those paths.  In observed data, however, comparisons across realized conversations generally compare
\[
Y_i(c) \quad \text{and} \quad Y_i(c'),
\]
across respondents for whom $C_i = c$ and $C_i = c'$.  Because these conversations are jointly produced, such comparisons reflect both the effect of the conversational path and differences in the respondents who helped generate those paths.  This implies that the realized conversation is both a treatment and a product of respondent characteristics that also affect outcomes.

\medskip\medskip

\noindent \emph{\textbf{Proposition 3 (Endogeneity of conversational paths).}
Comparisons of outcomes across realized conversations reflect both the causal effect of the conversational path and differences in respondent characteristics that influence the generation of those paths.}

\medskip\medskip

\noindent In other words, the realized conversation functions both as a treatment and as an indicator of respondent characteristics that also affect outcomes.\footnote{See Appendix~\ref{appendix_proof_prop3}.}

A further complication is that researchers often do not care about the full path $c$ as a sequence of ordered messages.  Instead, they care about features of the conversation.  As in previous sections, we can represent these features as functions of the conversational path, with $D_i = f(C_i)$ denoting observed features and $B_i = h(C_i)$ unobserved components.  The corresponding feature-level estimand is:
\[
\tau_D(d,d') = \E[Y_i(d) - Y_i(d')].
\]

\noindent This estimand is often closer to the theoretical object of interest than $\tau_C(c,c')$.  A researcher may not care about one transcript rather than another in all its details.  As in our anchoring example, they may instead care about a meaningful feature of the conversation---whether exposure to a more uncivil interaction changes their perception of democratic institutions as legitimate.

But moving from $C_i$ to $D_i$ still has identification challenges.  The measured feature $D_i$ is generated by the realized conversation, which is generated by both the interlocutor and the respondent (Figures~\ref{fig:conversation-dag} and~\ref{fig:conversation-dag-unpacked}).  Moreover, $D_i$ may be bundled with unmeasured features $B_i$ that \emph{also} affect outcomes.  A more uncivil conversation, for instance, may also be longer, more emotionally intense, more personally threatening, or more laden with identity.  Unless those other components are measured or controlled by design, differences in outcomes across values of $D_i$ cannot be attributed to the feature of interest alone \citep{Egami:2022ax,Fong:2023um}.

A limitation of the feature-based formulation is that it treats the conversational attribute as a static object.  In many settings, however, features vary over the course of the interaction.  A conversation may \emph{become} more or less civil, more or less informative, or more or less emotionally intense over time.  As a result, the relevant object may not be whether a feature is present in terms of the extensive margin, but how it accumulates throughout the conversation.  This naturally leads to a dosage interpretation of conversational features.  Let
\[
E_i = g(D_{i0},D_{i1},\ldots,D_{iT})
\]

\noindent denote cumulative exposure to a conversational feature over time.  Here $D_{it}$ represents the feature at each turn $t$, and $g(\cdot)$ aggregates those values into a measure of exposure.  For example, $E_i$ might capture the number of uncivil turns, the share of messages that are empathetic, or the cumulative amount of factual correction.  The corresponding dosage estimand is:
\[
\tau_E(e,e') = \E[Y_i(e) - Y_i(e')].
\]

\noindent This formulation extends the estimand based on features, allowing the effect of a conversational attribute to depend on things like its intensity or duration, rather than treating it as a binary or static aspect.  This object may be substantively useful because many theories of conversational influence are implicitly cumulative.  One uncivil sentence may not matter much, but repeated incivility might.\footnote{This dosage does not need to be monotonic either.  Trust might be increasing in empathy up to some threshold above which it becomes interpreted as forced or insincere, for example.  }  In this sense, dosage aggregates the dynamic evolution of conversational features into a single summary measure.  While this formulation captures a more realistic notion of conversational exposure, it does not, on its own, resolve the identification challenge.

\medskip\medskip

\noindent \emph{\textbf{Proposition 4 (Dosage inherits feature-level confounding).}
Comparisons of outcomes across levels of conversational exposure $E_i$ reflect both the effect of cumulative exposure to the feature of interest and the effect of other outcome-relevant components of the conversation that co-vary with that exposure.}

\medskip\medskip

\noindent This follows directly from the fact that $E_i$ is a function of the sequence of features $\{D_{it}\}$, so that any confounding present at the feature level is inherited by the aggregated object.\footnote{See Appendix~\ref{appendix_proof_prop4}}

These distinctions suggest that the effect of the full conversation, the effect of a conversational feature, and the effect of dosage are closely related but not identical.  The full-path estimand treats the transcript as the causal object.  The feature estimand treats some attribute of that transcript as the causal object.  While the dosage estimand treats accumulated exposure to that attribute as the causal object.  Each may be theoretically meaningful.  But their application depends on the question the researcher is asking.  Having established these estimands and the assumptions, the next section turns to more practical implications for research design.

\section{Implications for Research Design}

\noindent The framework developed above clarifies that conversational research does not involve a single causal object, but rather a set of related estimands corresponding to different interventions on the conversational process.  This has a straightforward first-order implication: researchers should decide which causal object they are interested in \emph{before} designing the study.

In practice, this decision is not always made explicitly, especially in the case where theoretical causes of interest are complex and researchers could be interested in different elements of them.  Researchers may randomize assignment to a conversational condition, but interpret their results as the effect of a message, a conversational feature, or the conversation itself.  The framework developed here suggests that this is not simply a matter of language.  These are different estimands; they answer different questions about the way that the world works; and moving between them requires careful attention to assumptions.

Crucially, we don't believe that retreating to only those estimands with easily defensible identification assumptions is the sole implication of our framework for research design.  Therefore, in this section, we walk through how research designs can be aligned with the causal objects of interest, and how alternative estimands can be defined when the ideal object is not directly manipulable.

\subsection{Match the estimand to the design}

\noindent The most basic design principle is to align the intervention with the causal object of interest \citep{Lundberg:2021ol}.  If the theoretical question concerns the effect of an opening message, then the opening message should be randomized and interpreted narrowly as such.  If the question concerns the effect of a conversational policy, then the policy should be randomized and the resulting effect interpreted as the effect of that regime.  In each case, the estimand will correspond to what is actually manipulated.\footnote{We focus on randomization as a means of addressing the relevant ignorability assumption.  But of course, other features of the design must be chosen with care in order to meet the other identification assumptions discussed in the relevant prior section.}

Problems arise when the intervention and the interpretation diverge.  For example, a researcher may assign respondents to an ``uncivil'' conversational condition, but then interpret differences in outcomes as the effect of exposure to uncivil messages or uncivil conversations.  As the earlier sections show, this interpretation generally requires additional assumptions, because the realized messages and conversations are not fixed by assignment alone.

\subsection{Move randomization closer to the conversational object of interest}\label{subsec:messagerandomization}

\noindent When the estimand is something beyond the coarse assignment object, then researchers can move randomization closer to the theoretical object of interest.  Rather than assigning respondents to a fixed conversational condition, the researcher may randomize aspects of the interaction.  For example, at a given conversational turn $t$, the researcher may observe the history $H_{it}$ and randomize the next message
\[
A_{it} \sim P(\cdot \mid H_{it}),
\]
where the distribution is chosen by design.  If one is interested in the effect of an uncivil \emph{message} within a conversation, rather than the effect of a generalized assignment or policy, this may be the effect of theoretical interest.

This approach is related to sequential multiple assignment randomized trials (SMART) and has two advantages \citep{Murphy:2005aa,Laber:2014ry}.\footnote{In that literature, individuals are randomized at multiple decision points, and later treatments may depend on information revealed by earlier treatment and response.  The goal is often to learn or compare adaptive treatment strategies.  Our proposed turn-level randomization uses the same design logic, but for a different inferential purpose.  Rather than primarily seeking an \emph{optimal} conversational regime, one can use sequential randomization to move the intervention closer to the conversational object of interest, such as a message or feature of a message at a particular history point.  Our framework emphasizes interpretation of causal effects more generally, staying agnostic about what is optimal or valuable.}  First, it reduces the extent to which variation in conversational features is driven by respondent-specific tendencies.  Part of the dependence between the respondent-generated history and the next message is broken at the randomized turn. Though we must still contend with the fact that the message \emph{meaning} may still be seen as a joint product of the history and the randomized message.  Second, it enables comparisons across alternative continuations of a given conversational state, or of histories that are treated as equivalent for the purposes of the design.

To be precise, the causal estimand identified by such a design is restricted to histories at which randomization occurs.  Let $Q_{it}$ denote an indicator that respondent $i$ is eligible for randomization at turn $t$ (i.e., the unit-history pair where the researcher intervenes).  Then the relevant estimand is
\[
\tau_{A_t}(a_t, a_t' \mid H_{it}, Q_{it}=1) = \E[Y_i(H_{it}, a_t) - Y_i(H_{it}, a_t') \mid Q_{it}=1].
\]

\noindent Substantively, this object captures the effect of alternative messages at specific points in the conversation where the researcher intervenes.  It therefore identifies a form of \emph{local} message effect, rather than the effect of messages across all possible conversational histories.\footnote{Here, $Q_{it}$ defines the subset of conversational states for which the researcher introduces variation.  It is one aspect fo the support of the estimand.}

Turn-level randomization induces a conditional ignorability condition by design.  Let $\mathcal{A}_{it}$ denote the set of admissible messages.  If the researcher randomizes $A_{it}$ conditional on the observed history $H_{it}$, then
\[
A_{it} \perp \{Y_i(H_{it},a): a \in \mathcal{A}_{it}\} \mid H_{it}, Q_{it}=1.
\]

\noindent Thus, among eligible histories, variation in the next message is generated by the researcher’s design rather than by the respondent or the policy alone. Still, this approach does not eliminate all challenges.  Histories remain high-dimensional, and the assumption that conditioning on $H_{it}$ removes all confounding remains strong.


A closely related approach is to relax the conditioning object itself.  Rather than conditioning on the full conversational history $H_{it}$, which may be too high-dimensional to be useful (as Section~\ref{sec:message-effect} discusses), the researcher may instead condition on a representation of that history; that is, a latent feature.  Let $S_{it} = \phi(H_{it})$ denote such a representation, which may capture features such as civility, topic, emotional tone, or a learned embedding of the prior interaction.  Importantly, this reflects a shift in the substantive question being asked.  Rather than asking about the effect of a message holding the full conversational history fixed---which, again, may be infeasible in practice---the researcher instead asks how the effect of a message varies across \emph{types} of conversational contexts defined by a representation of the history.

This leads to an estimands in the form of a Conditional Average Treatment Effect (CATE):
\[
\tau^{CATE}_{A_t}(a_t, a_t' \mid S_{it}) = \E[Y_i(S_{it}, a_t) - Y_i(S_{it}, a_t')].
\]

\noindent This estimand captures how the effect of a message depends on the conversational context in which it occurs.  For example, the same message have an effect in conversations that are already cooperative, but not have an effect in those that are hostile.\footnote{This links directly to the previous idea that the same message text may carry very different meanings across different conversational histories (e.g., Figure~\ref{fig:conversation-context}).  The factors that generate these interpretive differences are precisely the factors that must be accounted for in the conditioning set.  To the extent that they are unobserved or imperfectly captured, residual confounding remains.  One further complication arises when the relevance of past messages depends on timing or position within the interaction---for example, whether a feature appears in the opening message or later in the exchange.}  Rather than requiring exact matches on conversational histories, the researcher compares outcomes within groups of histories that are similar along dimensions of interest ($S_{it}$).  A practical extension is to discretize the representation---where representation is a lower-dimensional summary the conversational history---into coarse categories.  For example, histories may be grouped into low, medium, and high civility.  This can improve empirical support, but may also increase heterogeneity within groups, strengthening the required assumptions.\footnote{Appendix~\ref{appendix_proof_CATE} discusses assumptions related to the CATE.}

Once effects are allowed to vary across representations of conversational contexts, a natural question is how to summarize these heterogeneous effects.  In particular, the researcher must decide how to aggregate message effects across the distribution of conversational states observed in the data.  To capture this question, we can define the Average Marginal Component Effect of a message as
\[
\tau^{AMCE}_{A_t}(a_t, a_t') = \E_{S_{it}}[\tau^{CATE}_{A_t}(a_t, a_t' \mid S_{it})].
\]

\noindent This estimand averages message effects across representations $S_{it}$ of conversational contexts.\footnote{A reader familiar with conjoint analysis may wonder about the use of the term ``AMCE'' in this setting. We adopt the AMCE nomenclature because the logic is similar: a local effect is averaged over a distribution of contexts. Unlike standard conjoint designs, however, conversational contexts are not themselves randomly generated profile attributes. Rather, the weighting distribution arises from the observed distribution of conversational states (or the representations thereof) in the data. Thus, the analogy is not exact, but we feel like it reflects a common idea of averaging conditional effects across a distribution of contexts.}  It captures the average effect of a message across contexts in which it is observed, even if its effect varies across those different contexts.  In this sense, it summarizes the distribution of representational context-specific effects into a single quantity. Importantly, this aggregation is taken with respect to the distribution of conversational contexts in the data, so that the estimand reflects where conversations \emph{actually} occur. What this means is that identical context-specific effects may imply different AMCEs in populations that differ in the prevalence of conversational states.  Though related, the estimands are distinct---the CATE estimand captures context-specific effects, while the AMCE aggregates those effects across the distribution of conversational contexts.

These two different estimands can be illustrated using our anchoring example. In the case of the CATE, the researcher does not condition the message effect on the specific conversational history, but instead measures some meaningful features of it (e.g., civility, emotional intensity, and so forth).  The CATE then captures the effect of a civil as compared to uncivil interlocutor message conditional on a conversation that has had a certain degree of civility and emotional intensity up to that point.   This may be interesting to the researcher because is plausible that the effect of an uncivil message matters very differently in a conversation that has already been very uncivil, as compared to one that has not been, where it might be interpreted as jarring and offensive.  Rather than averaging over them as the message effect estimand does, the CATE distinguishes between them.

The AMCE, aggregates these heterogeneous effects. Suppose uncivil messages have strongly negative effects in highly civil conversations, modest negative effects in moderately civil conversations, and essentially no effect in highly uncivil conversations. The AMCE summarizes these context-specific effects into a single average quantity. As a result, the AMCE may indicate that uncivil messages are harmful on average, even though that average combines very different effects occurring in different conversational contexts in one study's observed data.

\subsection{Leverage strong policy as a source of variation}

\noindent Researchers can also think comprehensively about the AI policy as a source of variation in conversational exposure.  In particular, when a policy is a \emph{strong} lever---for example, when it assigns high probability to messages with a particular tone or feature---it can shift the distribution of conversational trajectories in ways that are less tightly coupled to respondent-specific tendencies.  Formally, a strong policy induces substantial variation in $A_{it}$ conditional on $H_{it}$, such that
\[
\Pr(A_{it} = a \mid H_{it}, \pi)
\]
\noindent differs meaningfully across values of $a$ even for similar histories.

Substantively, this corresponds to settings in which the conversational system exerts substantial influence over how the interaction unfolds.  This is most feasible in AI-mediated environments, where the researcher can shape the mapping from histories to messages.

This approach does not eliminate the underlying identification problem.  The policy remains a stochastic mapping from histories to messages, and the realized conversation is still jointly generated.  However, by inducing variation in conversational exposure that is less fully determined by respondent behavior, strong policies may improve the plausibility of identification assumptions for certain estimands.

\subsection{Constrained conversational protocols}

\noindent A distinct design strategy is to constrain the space of admissible messages within a conversation.  Rather than allowing the interlocutor or respondent to produce arbitrary text, the interaction is restricted to a predefined set of possible responses at each turn.

Formally, this corresponds to restricting the support of the message space.  At each turn $t$, the next message is drawn from a finite set
\[
A_{it} \in \mathcal{A}(H_{it}),
\]
\noindent where $\mathcal{A}(H_{it})$ is determined by the design.  In some cases, researchers might choose to set this as dependent on features of the history; in others, it may be fully fixed across conversations (e.g., \cite{Allamong:2025ffq}, \cite{Kline:2026yu}).

Substantively, this design reduces the dimensionality of the conversational process.  Instead of a high-dimensional and effectively unique history, the researcher observes repeated instances of similar conversational states and responses.  This can improve empirical support for comparisons, as the same or similar histories may reoccur with different messages.

Such designs are common in settings where interaction is structured.  For example, deliberative exercises may restrict participants to specific types of contributions, and conversational interfaces may provide suggested replies.  In these cases, the conversation is not fully free-form, but is instead guided by a protocol operating on the interlocutor and respondent that constrains what can be said.  However, one might worry that this approach trades realism for tractability.  If what defines a conversation is a certain degree of expressive freedom, then this restrictive protocol may destabilize the conceptual foundation on conversation.

It's also important to note that constraining the message space does not eliminate the core challenges identified above.  Even when messages are drawn from a finite set, their interpretation may depend on the conversational history.  The same response option may carry different meaning depending on the preceding interaction, and unobserved features of that history may still confound comparisons.  Thus, constrained conversational protocols provide a useful design tool for improving empirical tractability, but they do not fundamentally resolve the problem that conversational meaning is context-dependent and jointly generated.

\subsection{What can be recovered when the conversation cannot be assigned}

\noindent In many settings, the conversational object of theoretical interest---such as exposure to a particular type of interaction, tone, or feature---cannot be directly assigned by the researcher.  Assignment to a conversational condition may meaningfully influence the conversation, but does not fully determine it.  As a result, the realized conversational exposure may differ across respondents even within the same assignment condition.  Yet, researchers may still be interested in that realized conversation as a cause of certain outcomes.

One way to formalize what can be learned in such settings is by thinking of the realized conversation as evidence of compliance with an assignment condition.  When $Z_i$ denotes assignment to a conversational condition (opening message, or AI policy), and let $D_i = f(C_i)$ denote an observable feature of the realized conversation, such as its level of incivility, empathy, or intensity.  If assignment shifts the distribution of this feature without fully determining it, then $Z_i$ may be viewed as an instrument for $D_i$.

In this case, the estimand is no longer the average effect of conversational exposure in the population, but rather a local average treatment effect defined with respect to conversational features.  Let $D_i = f(C_i)$ denote a feature of the realized conversation, and let $D_i(z)$ denote the feature that would be realized under assignment $z$.  Then the relevant estimand is:
\[
\tau^{LATE}_D(d, d') = \E\big[ Y_i(d) - Y_i(d') \mid D_i(z) = d,\; D_i(z') = d' \big],
\]
\noindent which captures the effect of conversational exposure among respondents whose realized conversations are shifted by the assignment.

Substantively, this corresponds to a different question from those considered above.  Rather than asking what would happen if all respondents experienced a particular conversational feature, it asks: what is the effect of that feature among those respondents whose conversational trajectories are shifted by the assigned condition?  Some people comply with a conversational condition, while others don't.  For example, if assignment to an uncivil conversational policy increases the likelihood that some respondents experience uncivil interaction, then the corresponding LATE captures the effect of uncivil interaction \emph{for those respondents} whose conversations are made more uncivil by that assignment.

Interpreting this estimand requires additional assumptions (see Appendix~\ref{appendix_proof_LATE} for a full discussion).  First, assignment must affect conversational exposure (relevance).\footnote{In practice, manipulation-check questions concerning perceived civility, hostility, empathy, or persuasion may provide one way of operationalizing conversational uptake, though such measures do not eliminate the substantive assumptions underlying a LATE interpretation.}  Second, assignment must affect the outcome only through the conversational feature of interest (exclusion).  Third, the direction of the effect of assignment on exposure must be consistent across respondents (monotonicity).  Finally, assignment must be independent of potential outcomes (as in Assumption 3 above).

These assumptions are still demanding in conversational settings.  In particular, consider the exclusion restriction.  We may theorize that assignment influences outcomes through multiple aspects of the interaction, including the opening message, the length of the conversation, the number of turns taken, and specific latent features like the overall degree of incivility.  Without a sufficient number of instruments for those features, exclusion may be violated.

As another example, monotonicity may be violated if the researcher attempts to induce engagement (the theoretical causal factor of interest) with a certain treatment, but respondents react in opposing ways (some comply and some defy)---for example, some respondents may become more engaged with incivility while others disengage.  This suggests that understanding, in a narrow sense, how respondents experience a treatment that will be used as an instrument is an important pre-condition for this kind of estimand.

The value of this approach is therefore not that it recovers the full effect of the conversational object of interest, but that it provides a principled way of linking assignment to realized conversational exposure when direct manipulation is not possible.  In this sense, it offers a complementary strategy to the approaches described above: rather than moving the point of randomization closer to the conversational object, it uses the existing assignment to recover a well-defined, if distinct, causal effect.

\section{Extending to Human Interlocutors}\label{sec:humans}

\noindent The framework developed above treats the interlocutor’s behavior as governed by a conversational policy.  This is a natural representation for AI-mediated interactions, where the mapping from conversational histories to messages is determined by a system that is ``known'' in certain respects.  In many settings of interest for scholars of politics, however, the interlocutor is a human agent, and variation in conversational behavior reflects heterogeneity across individuals rather than a single underlying policy \citep{Pons:2018vv,Broockman:2016mq,Kalla:2020oc}.  Even if we assume random assignment of respondents to interlocutors, this representation introduces several additional inferential concerns that we highlight here.

First, the conversational policy is no longer a single object induced by assignment, but is instead bundled with the interlocutor.  Each interlocutor brings their own tendencies---in tone, responsiveness, patience, aggressiveness, and interpretation of prior messages---so that $\pi_j$ reflects both the intended design and idiosyncratic behavior \citep{Zhang:2020qp}.  At the same time, interlocutors themselves are not passive conduits of the treatment.  Their behavior is shaped by their own latent characteristics, prior experiences, and situational judgments.  As a result, the policy is not merely assigned; it is likely enacted differently across individuals, and this heterogeneity becomes part of the causal process.

Second, the meaning of a message is not separable from identity of the interlocutor in quite the same way.  The same text may be interpreted differently depending on who says it and to whom it is said.\footnote{We note that this question of interpretation is a deep one that exists at the intersection between social identities, meaning-making and causality. }  A conciliatory statement from a trusted peer may be received as genuine, while the same statement from an out-group member may be dismissed or viewed with suspicion.  Attempting to hold fixed relevant interlocutor identities may be challenging, and those that are not explicitly presented, may be inferred in systematic but unmeasurable ways that affect outcomes \citep{Flekova:2016dc}. Human interlocutors highlight even more strongly the relational nature of message: their causal effect depends not only on their content and the prior conversational history, but also on the identify of, and social relationship between interlocutor and respondent.

Third are issues of spillovers.  As noted before, AI policy can feedback on itself in ways that generate spillovers between respondents if the AI is allowed to learn from its conversations.  These issues are likely to be exacerbated in human-interlocutor conversations.  Human interlocutors often participate in multiple conversations, and their behavior in one interaction may influence their behavior in subsequent ones.  In addition to learning; fatigue, strategic adaptation, or emotional carryover can all create dependencies across conversations, violating no interference assumption.

Finally, the treatment itself may extend beyond the textual content of messages.  When human-mediated conversations occur face-to-face, or beyond text, aspects such as tone of voice, timing, facial expression, and other non-verbal cues may influence both the course of the conversation and the outcome.  These paralinguistic elements are often unobserved or difficult to measure, yet they may be central to how messages are interpreted \citep{Wang:2017dd,VanZant:2020ps}.  As a result, even a complete transcript may not fully capture the relevant treatment as experienced.

Taken together, these considerations suggest that extending the framework to human interlocutors does not simply add noise to an otherwise well-defined process.  Various aspects of the treatment change.  The conversational policy becomes intertwined with interlocutor identity; message meanings transform; and the treatment may extend beyond what is observed in the transcript.  We suggest that these factors require additional care in defining estimands and interpreting results.

\section{Conclusion}

\noindent Conversations are not fixed treatments.  They are sequential, interactive, and jointly generated processes.  This simple observation has important consequences for causal inference and the link between theory and empirics in the conversational domain.  When treatments take the form of conversations, the object that is assigned by the researcher only partly shapes what is experienced by the respondent.  As a result, causal claims about conversational content, features, or trajectories require greater care than in settings with static interventions.

This paper develops a framework for making those distinctions explicit.  We define a set of causal estimands corresponding to different objects that arise in conversational settings---assignment, opening messages, conversational policy, messages within a conversation, and the realized conversational process itself, including features of it (e.g., length, content, tone).  These objects are not interchangeable.  Each corresponds to a different substantive question, distinct theoretical object, and therefore a different implied counterfactual.  Thus, there are different assumptions required for each of them in order to make causal inferences.  In particular, while assignment and policy effects may be identified under standard experimental designs, message-level and feature-level effects generally are not, due to the joint generation of conversational histories and the bundling of outcome-relevant components within messages.

The contribution of the paper is \emph{not} to suggest that researchers should restrict attention to the most easily identified estimands.  Rather, it is to clarify the tradeoffs involved in studying causal objects that more closely accord with rich theory of how and why the political world operates as it does---not merely what policies we can implement to make change in it.  Researchers are unlikely to stop reasoning about the effects of conversational content, tone, or interaction patterns.  This paper makes clear how doing so requires certain assumptions, research design choices, or particular estimands that are explicitly tied to the structure of conversational data.  Making those assumptions visible allows scholars to better align their empirical strategies with their theoretical goals, interpret their findings, and, importantly, to innovate in those strategies.

More broadly, the framework speaks to how to approach causal inference in dynamic settings with jointly-produced content.  As treatments become more realistic and responsive---particularly with the increasing use of AI-mediated conversations---the gap between what can be assigned and what is ultimately experienced may widen.  This is not a simple ``failure'' of causal inference, but a reflection of the fundamental complexity of the social processes under study.  Conversations are powerful precisely because they are adaptive and co-produced.  Those same features, however, make causal effects harder to define and identify.

Several avenues for future research follow from this work.  First, there is scope for further developing experimental designs that move the point of intervention closer to the conversational objects of interest, potentially incorporating dynamically measured outcomes and conversational expectation.  Second, further work is needed to better understand how to represent conversational histories in ways that balance what makes those histories theoretically rich, while also allowing them to be empirically tractable.  Third, as we've noted, extending the framework to settings with human interlocutors raises additional issues that must be dealt with carefully.

\singlespacing
\clearpage
\newpage

\bibliographystyle{apsr.bst}
\bibliography{/Users/Vesper/Documents/Bibtex/Complete_Library.bib}

\clearpage
\newpage
\setcounter{section}{0}
\renewcommand{\thesection}{\Alph{section}}
\renewcommand{\thesubsection}{\thesection.\arabic{subsection}}
\section{Appendix: Decompositions and Formalizations}\label{appendix:proofs}

\subsection{Formalization of Proposition 1: Selection Bias from Endogenous Histories}\label{appendix_proof_prop1}

\noindent This appendix shows a point made throughout the main text of the paper: messages within conversations are not assigned independently of the conversational process that generates them. Instead, they come about from histories that are jointly produced by the respondent and the interlocutor. This appendix shows how this endogeneity of conversational histories induces selection bias when researchers attempt to estimate message effects.

For notational simplicity, we suppress the time index $t$ and write $A_i$ for the message, $H_i$ for the conversational history that precedes that message, and $Y_i(h,a)$ for the potential outcome in the state of the world where the conversational history is $h$ and the message is $a$.  The causal object of interest is the effect of changing the message while holding fixed the conversational history:
\[
\tau_A(a,a' \mid H_i=h)
=
\E\!\left[
Y_i(h,a)
-
Y_i(h,a')
\mid H_i=h
\right].
\]

\noindent This quantity compares two states of the world that differ only in the message received, while holding fixed everything that has occurred up to that point in the conversation.  However, in terms of what we observe, we do not observe both potential outcomes for the same respondent. Instead, observationally, we compare respondents who happen to receive different messages:
\[
\hat{\tau}_A(h)
=
\E[Y_i \mid A_i=a,H_i=h]
-
\E[Y_i \mid A_i=a',H_i=h].
\]

\noindent By consistency, $Y_i = Y_i(H_i,A_i)$, we can write that
\[
\hat{\tau}_A(h)
=
\E[Y_i(h,a)\mid A_i=a,H_i=h]
-
\E[Y_i(h,a')\mid A_i=a',H_i=h].
\]

\noindent We can now compare what we observe to what we would ideally like to observe. To do that we can define our bias term as
\[
Bias(h)
=
\hat{\tau}_A(h)
-
\tau_A(a,a' \mid H_i=h).
\]

\noindent Substituting the definitions above and rearranging yields
\[
\begin{aligned}
Bias(h)
&=
\Big(
\E[Y_i(h,a)\mid A_i=a,H_i=h]
-
\E[Y_i(h,a)\mid H_i=h]
\Big)
\\
&\quad
-
\Big(
\E[Y_i(h,a')\mid A_i=a',H_i=h]
-
\E[Y_i(h,a')\mid H_i=h]
\Big).
\end{aligned}
\]

\noindent This decomposition shows that the difference we observe equals the desired causal effect only when conditioning on the realized message provides no additional information about potential outcomes beyond what is already contained in the history.

This motivates the question as to why these additional terms may fail to equal zero.  In order to see this, suppose there exists an unobserved respondent characteristic $U_i$ that affects both conversational behavior and outcomes as in the DAGs presented in the main text. Examples include generalized traits and knowledge like political sophistication, curiosity, hostility, verbosity, resistance to persuasion, or willingness to engage with disagreement.  These respondent characteristics influence the messages that the respondent contributes to the conversation, which in turn influence the conversational history (by how the interlocutor and the respondent themselves subsequently produce messages).

Consequently, the observed message contains information about respondent characteristics that also affect outcomes. Even after conditioning on the realized conversational history, any aspects of respondent behavior that are not \emph{fully captured} by $H_i$ remain associated with both the message and the outcome.
\[
\Pr(A_i=a \mid H_i=h,U_i=u)
\neq
\Pr(A_i=a \mid H_i=h),
\]

\noindent while

\[
\E[Y_i(h,a)\mid H_i=h,U_i=u]
\]

\noindent may vary across values of $u$.  This means that respondents who receive message $a$ need not be comparable to respondents who receive message $a'$, even when they share the same observed conversational history. The observed contrast therefore combines the causal effect of the message with differences in respondent characteristics that helped generate the conversational path leading to that message.

The key implication is that the source of bias is not merely the failure of sequential ignorability as an abstract condition. But instead we can think of sequential ignorability failing because conversational histories are endogenous products of interaction. Respondents are not passive recipients of treatment. They help generate the histories that determine which messages are observed.  Because of that, Proposition 1 does not require that histories be completely unobserved, only that histories are generated by respondent characteristics.

\medskip \medskip
\subsection{Formalization of Proposition 2: Feature Bundling Bias}\label{appendix_proof_prop2}

\noindent This appendix shows that even if a researcher could isolate a message from the conversational process that generated it, the message itself may contain multiple outcome-relevant features. As a result, variation in a feature of interest may remain bundled together with other aspects of the message that also affect outcomes.

For simplicity, let $D_i$ denote a measured feature of a message, such as incivility, empathy, or factual correction, and let $B_i$ denote all other outcome-relevant features of that same message that are not included in the analysis.  And we'll assume that outcomes depend on both $Y_i = Y_i(D_i,B_i)$.

The causal object of interest is the effect of changing the feature $D_i$ while holding fixed all other aspects of the message:
\[
\tau_D(d,d')
=
\E\!\left[
Y_i(d,B_i)
-
Y_i(d',B_i)
\right].
\]

\noindent This quantity compares two states of the world that differ only in the feature of interest.  In practice, however, researchers compare outcomes across observed values of $D_i$:
\[
\hat{\tau}_D(d,d')
=
\E[Y_i \mid D_i=d]
-
\E[Y_i \mid D_i=d'].
\]

\noindent By consistency, we have that
\[
Y_i
=
Y_i(D_i,B_i),
\]

\noindent and then we can write
\[
\hat{\tau}_D(d,d')
=
\E[Y_i(d,B_i)\mid D_i=d]
-
\E[Y_i(d',B_i)\mid D_i=d'].
\]

\medskip

\noindent We can again compare what we observe to what we would like to observe by defining
\[
Bias(d)
=
\hat{\tau}_D(d,d')
-
\tau_D(d,d').
\]

\noindent Adding and subtracting the relevant terms yields
\[
\begin{aligned}
Bias(d)
&=
\Big(
\E[Y_i(d,B_i)\mid D_i=d]
-
\E[Y_i(d,B_i)]
\Big)
\\
&\quad
-
\Big(
\E[Y_i(d',B_i)\mid D_i=d']
-
\E[Y_i(d',B_i)]
\Big).
\end{aligned}
\]

\noindent This expression shows that the observed contrast equals the desired causal effect only if the distribution of bundled features $B_i$ is independent of the feature of interest.  A sufficient condition for the observed contrast to equal the desired causal effect is that the bundled features be independent of the feature of interest:
\[
D_i \perp B_i.
\]

\noindent Because both $D_i$ and $B_i$ are functions of the same message,
\[
A_i \rightarrow (D_i,B_i),
\]

\noindent independence is generally not expected to hold without additional design choices or assumptions.  When this condition fails, changes in the feature of interest are accompanied by changes in other outcome-relevant aspects of the message.

For example, messages that differ in incivility may also differ in emotional intensity, length, specificity, identity signaling, or argument structure. Consequently, respondents exposed to different values of $D_i$ are simultaneously exposed to different values of $B_i$.  The observed contrast therefore combines the causal effect of the feature of interest with the effect of other outcome-relevant components that vary alongside it.

This is the concern emphasized throughout the main text. Messages are bundles. Even if a feature can be measured, variation in that feature does not generally occur in isolation. As a result, comparisons across values of a feature recover the effect of features and bias term due to bunding, where the bundling bias arises because messages contain multiple outcome-relevant components that covary with the feature of interest.

\medskip \medskip
\subsection{Comment on Proposition 3: Endogeneity of Conversational Paths}\label{appendix_proof_prop3}

\noindent Proposition 3 follows directly from the logic developed in Proposition 1 (see Appendix~\ref{appendix_proof_prop1}). In that case, the object of interest was a message generated from an endogenous conversational history. Here, the object of interest is the full realized conversation itself.

The key insight is unchanged. Respondent characteristics $U_i$ help generate the conversational process while also affecting outcomes. Consequently,
\[
Y_i \leftarrow  U_i \rightarrow C_i.
\]

\noindent And so, as a result, realized conversations need not be independent of potential outcomes. The observed difference
\[
\E[Y_i \mid C_i=c]
-
\E[Y_i \mid C_i=c']
\]

\noindent combines the causal effect of the conversation with selection induced by the conversational process itself.  The derivation is formally identical to that in Proposition 1, replacing the message object $A_i$ with the realized conversational path $C_i$. For this reason we do not repeat the full decomposition here.

\medskip \medskip
\subsection{Comment on Proposition 4: Dosage Inherits Feature-Level Confounding}\label{appendix_proof_prop4}

\noindent Proposition 4 follows directly from the logic developed in Proposition 2 (Appendix~\ref{appendix_proof_prop2}). In that case, the object of interest was a conversational feature $D_i$. Here, the object of interest is a dosage measure
\[
E_i
=
g(D_{i1},D_{i2},\ldots,D_{iT}),
\]

which is constructed by aggregating features across the course of a conversation.  The key insight is unchanged. If variation in the underlying features $D_{it}$ remains bundled together with unmeasured outcome-relevant components, then any function of those features inherits the same dependence structure. Consequently, comparisons across dosage levels do not isolate the effect of cumulative exposure alone.

In this sense, dosage does not resolve the feature-bundling problem. It aggregates it. The observed contrast therefore combines the effect of cumulative exposure with the effects of other conversational components that covary with that exposure.

The formal derivation is identical to Proposition 2 after replacing the feature object $D_i$ with the dosage object $E_i=g(D_{i1},\ldots,D_{iT})$. For this reason we do not repeat the full decomposition here.

\clearpage
\newpage
\section{Appendix: Alternative Estimands}\label{appendix:altest}

\medskip \medskip
\subsection{Identification Decomposition in Representation Sufficiency for Message-Level CATEs}
\label{appendix_proof_CATE}

\noindent This appendix makes explicit a point that is discussed in the main text: conditioning on a representation of conversational history, rather than the full history itself, can introduce bias.  The goal of the proof is to show exactly where that bias comes from.

In the main text, we allow researchers to replace the full conversational history $H_{it}$ with a lower-dimensional representation.  They may do this to address the dimensionality of conversational histories. The lower level representation is:
\[
S_{it} = \phi(H_{it}).
\]

\noindent The hope for the researcher is that $S_{it}$ captures the relevant aspects of the history for both message generation and outcomes, while avoiding the problem that the full history may be sparse, making conditioning difficult in practice.

For notational simplicity, we wills et aside the time index $t$, and write $A_i$ for the message, $S_i$ for the representation, and $Y_i(s,a)$ for the potential outcome in the state of the world where the message is $a$ when the represented history is $s$.  The causal object of interest is the Conditional Average Treatment Effect comparing $a$ and $a'$ under condition $s$:
\[
\tau^{CATE}_{A}(a,a' \mid S_i=s)
=
\E\left[
Y_i(s,a) - Y_i(s,a')
\mid S_i=s
\right].
\]

\noindent This is the object we would like to learn: the effect of changing the message while holding fixed the represented conversational state.  In the data, however, we do not observe both potential outcomes for the same unit.  Instead, we compare units who receive different messages.  This makes the observed contrast among units with $S_i=s$ is:
\[
\hat{\tau}^{CATE}_{A}(s)
=
\E[Y_i \mid A_i=a, S_i=s]
-
\E[Y_i \mid A_i=a', S_i=s].
\]

\noindent Which we can expand, via our consistency assumption, because when $A_i=a$ we observe $Y_i=Y_i(s,a)$, and when $A_i=a'$ we observe $Y_i=Y_i(s,a')$.  Substituting, we obtain:
\[
\hat{\tau}^{CATE}_{A}(s)
=
\E[Y_i(s,a) \mid A_i=a, S_i=s]
-
\E[Y_i(s,a') \mid A_i=a', S_i=s].
\]

\medskip

\noindent We can then compare what we observe to what we would like to observe, and then rearrange to see the bias term.  The estimand conditions only on $S_i=s$, while the observed contrast conditions on both $S_i=s$ \emph{and} the realized message.  The difference between these two objects is the source of bias:
\[
\begin{aligned}
Bias(s)
&=
\hat{\tau}^{CATE}_{A}(s)
-
\tau^{CATE}_{A}(a,a' \mid S_i=s).
\end{aligned}
\]

\noindent Adding and subtracting terms, we can write:
\[
\begin{aligned}
Bias(s)
&=
(
\E[Y_i(s,a) \mid A_i=a, S_i=s]
-
\E[Y_i(s,a) \mid S_i=s]
) \\
&\quad -
(
\E[Y_i(s,a') \mid A_i=a', S_i=s]
-
\E[Y_i(s,a') \mid S_i=s]
).
\end{aligned}
\]

\medskip

\noindent This expression makes the problem clear.  The observed contrast will equal the desired causal effect if and only if these two quantities are equal:
\noindent The expression
\[
\E[Y_i(s,a) \mid A_i=a, S_i=s]
=
\E[Y_i(s,a) \mid S_i=s]
\]

\noindent states that, among units with represented history $s$, those who happen to receive message $a$ are not systematically different from those who do not, in terms of their potential outcomes for $a$.  That is, conditioning on the realized message does not change the expected outcome in the state of the world where the message is set to $a$.  This is equivalent to a form of ignorability at the level of the history representation:
\[
A_i \perp Y_i(s,a) \mid S_i=s.
\]

\noindent  This can be thought of as a kind of \emph{representation sufficiency}---the representation must capture all aspects of the history that jointly influence both message assignment and outcomes.

The key question, then, is what happens when this condition fails.  To see this more concretely, suppose there exists an unobserved respondent characteristic $U_i$ that affects both the message that is produced and the outcome.  In that case:
\[
\Pr(A_i=a \mid S_i=s, U_i=u)
\neq
\Pr(A_i=a \mid S_i=s),
\]

\noindent  so that the probability of observing a certain message depend on $U_i$ \emph{even after} conditioning on $S_i=s$.  At the same time potential outcomes for a message and history representation:
\[
\E[Y_i(s,a) \mid S_i=s, U_i=u]
\]

\noindent  may vary across values of $u$.  That is, different types of respondents have different potential outcomes for the same message, even within the same represented history.  Consequently, the distribution of $U_i$ differs across message conditions.
And so, within the same represented history, the treated and control groups differ in their composition of respondent types.  The observed contrast therefore combines the causal effect of the message with differences in the types of respondents who receive each message.

This is precisely the concern emphasized in the main text.  The representation $S_i$ may summarize the observable conversational history, but it need not capture all aspects of the interaction that matter for both message generation and outcomes.  When it fails to do so, comparisons within $S_i=s$ do not recover the desired causal effect.

\medskip \medskip
\subsection{Identification Assumptions for Conversational LATEs}
\label{appendix_proof_LATE}

\noindent This appendix develops a point discussed in the main text---assignment to a conversational condition may sometimes be used as an instrument for a conversational feature of interest. The appeal of this approach is that the feature itself may not be directly manipulable or randomly assigned, even though assignment to a conversational condition is.  Unlike the previous decompositions, the LATE framework does not attempt to eliminate endogeneity of conversational features directly. Instead, it seeks to leverage experimentally induced variation in those features through assignment, and shift the estimand to something with more plausibly met assumptions (though still difficult).

Recall that the object of interest is a conversational feature $D_i=f(C_i)$ generated by the realized conversation. Examples might include the level of incivility, empathy, perspective-taking, factual correction, or some other characteristic of the interaction. Assignment $Z_i$ influences the conversation, which in turn influences the feature:
\[
Z_i \rightarrow C_i \rightarrow D_i.
\]

\noindent The Local Average Treatment Effect (LATE) considered in the main text is

\[
\tau_D^{LATE}
=
\E
\left[
Y_i(d)-Y_i(d')
\;\middle|\;
D_i(z)=d,\;
D_i(z')=d'
\right].
\]

\noindent This quantity captures the effect of the conversational feature among respondents whose conversational exposure changes as a consequence of assignment. The appeal of this estimand is that it moves the analysis closer to the conversational object of theoretical interest than the assignment effect alone. The key question is what assumptions are required for such an interpretation.

First, assignment must affect the conversational feature. Formally,
\[
D_i(z)
\neq
D_i(z')
\]

\noindent for at least some respondents. This is the familiar relevance condition in the context of instrumental variables. If assignment does not shift the conversational feature, then it cannot be used to learn about the effects of that feature.

Second, assignment must affect outcomes only through the conversational feature of interest. This is the exclusion restriction. In conversational settings, however, this assumption is often difficult to justify because assignment may influence \emph{multiple} aspects of the interaction at the same time.

To see this, suppose assignment affects both a measured conversational feature $D_i$ and other outcome-relevant conversational characteristics $B_i$ such that:
\[
Z_i
\rightarrow
(D_i,B_i)
\rightarrow
Y_i.
\]

\noindent In that case, differences in outcomes induced by assignment reflect not only the effect of the feature of interest, but also the effects of \emph{other} conversational changes generated by the intervention. The exclusion restriction therefore requires that assignment affect outcomes through $D_i$ alone.

This assumption may be particularly demanding in conversational settings. An intervention designed to increase empathy, for example, may also alter conversational length, engagement, emotional intensity, or trust. Likewise, an intervention intended to increase factual correction may simultaneously alter tone, and something about the perception of legitimacy or expertise.

Third, monotonicity requires that assignment shift the conversational feature in a common direction:
\[
D_i(z)
\geq
D_i(z')
\text{ \hspace{.1in} or \hspace{.1in} }
D_i(z)
\leq
D_i(z')
\]

\noindent for all respondents.  There cannot be those who defy the treatment as induced by the instrument.\footnote{Or there can be defiers but in that case there can be no compliers.}  In conversational settings this assumption may be less obvious than in many conventional instrumental variable applications. The same assignment may increase exposure to a feature for some respondents while decreasing it for others. For example, assignment to an adversarial conversational policy may increase hostility among some respondents but cause others to actively reframe the conversation as congenial, or to disengage entirely. In such cases, the interpretation of the resulting LATE becomes more difficult.

Finally, assignment must be independent of potential outcomes. In experimental settings this condition is often justified by design. However, the previous appendices demonstrate that the conversational features generated after assignment are not themselves randomized objects. The value of the instrumental variable approach is therefore not that it eliminates these concerns, but that it shifts identification onto assumptions about how assignment affects conversational features.

Taken together, these conditions illustrate both the appeal and the limitation of conversational LATEs. They provide a principled way of moving from assignment effects toward conversational features that may be of greater theoretical interest. At the same time, they do so by replacing one set of assumptions with another. In particular, the exclusion restriction requires that assignment influence outcomes only through the feature of interest, a condition that may be especially challenging to justify when conversational interventions alter multiple aspects of an interaction simultaneously.

\end{document}